\documentclass[preprint, twocolumn,showpacs,showkeys,10pt, amsmath,amssymb, aps, prc, floatfix]{revtex4-2}

\usepackage{graphicx}
\usepackage{dcolumn}
\usepackage{bm}
\usepackage {xcolor}
\usepackage[colorlinks = true,linkcolor = blue,urlcolor  = blue,citecolor = blue,anchorcolor = blue]{hyperref}
\usepackage{times}
\usepackage{float}

\begin{document}

\title{Twin stars as probes of the nuclear equation of state: effects of rotation through the PSR J0952-0607 pulsar and constraints via the tidal deformability from the GW170817 event}

\author{Lazaros Tsaloukidis}
\email{ltsalouk@pks.mpg.de}
\affiliation{Max Planck Institute for the Physics of Complex Systems (MPI-PKS)}
\affiliation{W\"urzburg-Dresden Cluster of Excellence ct.qmat. 01187 Dresden, Germany}
\author{P.S. Koliogiannis}%
 \email{pkoliogi@physics.auth.gr}
\author{A. Kanakis-Pegios}
\email{alkanaki@auth.gr}
\author{Ch.C. Moustakidis}
\email{moustaki@auth.gr}
 \affiliation{Department of Theoretical Physics, Aristotle University of Thessaloniki, 54124 Thessaloniki, Greece
}%

\date{\today}

\begin{abstract}
In agreement with the constantly increasing gravitational wave events, new aspects of the internal structure of compact stars can be considered. A scenario in which a first order transition takes place inside these stars is of particular interest as it can lead, under conditions, to a third gravitationally stable branch (besides white dwarfs and neutron stars), the twin stars. The new branch yields stars with the same mass as normal compact stars but quite different radii. In the present work, we focus on hybrid stars undergone a hadron to quark phase transition near their core and how this new stable configuration arises. Emphasis is to be given on the aspects of the phase transition and its parametrization in two different ways, namely with Maxwell and Gibbs construction. We systematically study the gravitational mass, the radius, and the tidal deformability, and we compare them with the predictions of the recent observation by LIGO/VIRGO collaboration, the GW170817 event, along with the mass and radius limits, suggesting possible robust constraints.  Moreover, we extent the study in order to include rotation effects  on the twin stars configurations. The recent discovery of the fast rotating supermassive pulsar PSR J0952-0607 triggered the effort to constrain the equation of state and moreover to examine possible predictions related to the phase transition in dense nuclear matter. In the present work we pay special attention to relate the  PSR J0952-0607 pulsar properties with the twin stars predictions and mainly to explore the possibility that the existence of such a massive object would rule out the existence of twin stars. Finally, we discuss the constraints on the radius and mass of the recently observed compact object within the supernova remnant HESS J1731-347. The estimations imply that this object is either the lightest neutron star known, or a  star  with a more exotic equation of state.
\end{abstract}

\maketitle

\section{Introduction}
Compact stars yield the most prominent natural laboratories for the study of exotic forms of matter~\cite{Glendenning-1997,Haensel-2007,Bielich-2020,Weber-1999}. Recently discovered pulsars alongside with gravitational waves detection, such as GW170817, have revealed new aspects of the internal structure of these stars, mainly in terms of their composition~\cite{Abbott-1,Abbott-2,Abbott-gw170817}. Whilst the equation of state (EOS) of nuclear matter is well established up to nuclear saturation density, one encounters the challenge of describing matter in fairly higher densities realized in the interior of these stars. At these densities the type of matter is yet to be determined and in turn, the construction of stellar models to agree with the aforementioned observations is still an open issue. Possible candidates are pure neutron stars composed by hadrons, strange quark stars composed of deconfined quarks,  and hybrid stars composed by hadronic outer shells and cores of deconfined quarks. In the present article the latter case is claimed.

Stars of this branch are expected to have masses in the same range as normal neutron stars, yet fairly smaller radii. The existence of such stars is a strong indication that a hadron-quark phase transition (HQPT) is a physical reality, a result of utmost importance, especially in the study of dense matter physics~\cite{Baym-2018,Alford-2013,Christian-2018,Montana-2019}. A study of the HQPT is presented here, examining the conditions under which the twin star configuration arises. The compatibility with the mass and radius constraints, as they are formed through up to date observations, is also considered.

The idea of a third family of compact stars and in particular, the connection with the possibility to be a signature of a strong phase transition in the interior of the star, was firstly introduced by  Gerlach~\cite{Gerlach-1968}. Later on, K{\"a}mpfer worked also on this issue~\cite{Kampfer-1981a,Kamfer-1981b}. Glendenning and Kettner introduced the term ``twins" in their paper~\cite{Glendenning-2000}, while at the same time Schertler {\it et al.}~\cite{Schertler-2000} worked out that idea in detail. However, in all the previous studies, the maximum mass was approximately at the canonical binary pulsar mass $1.4~M_{\odot}$. The revival of the idea of the twin stars started a few years later by Blaschke {\it et al.}~\cite{Blaschke-2013,Castillo-2013}. Specifically, in the mentioned papers is suggested that high-mass twin stars, once detected by simultaneous mass and radius measurements, could provide the evidence for a strong first-order phase transition in cold matter, which then would imply the existence of at least one critical endpoint in the quantum chromodynamics phase diagram. Moreover, in the same work are also presented, for the first time, examples of EOSs that would not only provide twin solutions, but also fulfill the constraint on the maximum mass from the existence of pulsars as heavy as $2~M_{\odot}$~\cite{Blaschke-2013,Castillo-2013}. The previous idea was elaborated by Benic {\it et al.}~\cite{Benic-2015} (see also Ref.~\cite{Castillo-2015}). A systematic Bayesian analysis of the new twin star EOS with observational constraints was presented in Ref.~\cite{Castillo-2016}. Finally, the analysis of the robustness of twin solutions against changing the Maxwell to a mixed phase construction, and the formation of structures in the mixed phase due to the interplay of the surface tension and Coulomb interaction effects, have been also considered respectively in Refs.~\cite{Ayriyan-2018,Maslov-2019}. Some recent works dedicated to the study of phase transition in the interior of the neutron stars and possible existence of  twin stars, are included in  Refs.~\cite{Alvarez-2017,Bejger-2017,Bhattacharyya-2010,Christian-2019,Christian-2021,Christian-2022,Han-2019a,Espinoza-2022,Tan-2022,Li-2021,Sharifi-2021,Benitez-2021,Sendra-2020,Castillo-2019,Paschalidis-2018,Spinella-2017,Alford-2017,Zacchi-2017,Zacchi-2016,Bhattacharyya-2005,Sen-2022,Minamikawa-2021,Pietri-2019,Zdunik-2013,Han-2020,Li-2020,Largani-2022,Ivanytskyi-2022,Contrera-2022,Schram-2016,Burgio-2018,Sieniawska-2019,Most-2018,Nandi-2018,Deloudis-2021,Wang-2022,Banik-2004,Banik-2001,Banik-2003,Haensel-2016,Bozzola-2019,Pereira-2020,Han-2019}.  

Pulsar PSR J0952-0607 was firstly discovered by Bassa {\it et al.}~\cite{Bassa-2017} with a frequency of $f=709$ Hz, making it the fastest known spinning pulsar in the disk of Milky Way. Very recently, Romani {\it et al.}~\cite{Romani-2022} discovered that PSR J0952-0607 has a mass of $M=2.35\pm 0.17 \ M_{\odot}$, the largest well measured mass found up to date. The above discoveries triggered the research into the constraints of the EOS of dense nuclear matter and is likely to revise many of the theoretical predictions concerning basic properties of neutron stars (for a recent study see Ref.~\cite{Ecker-2022}). One of them is the existence of twin stars, which is the main subject of the current study. In particular, the existence of a supermassive neutron star may rule out twin stars (see Ref.~\cite{Christian-2021}). In this sense, the observation of such supermassive stars is decisive for checking the reliability of the corresponding theoretical predictions and therefore, a systematically and careful analysis should be consider. On the other hand, theoretical calculations for the maximum possible mass of neutron stars should be combined with recent observations concerning the radius and tidal deformability of stars with mass close to $1.4~M_{\odot}$. In general, a nuclear model can be judged by its flexibility in being able to predict the maximum possible mass while correctly estimating the observational radial and tidal deformability values.

Moreover, one of the motivations of the present work is to examine in a more systematic way the applications of the two main formulation of the phase transition, that is the Maxwell (MC) and Gibbs (GC) constructions. These formulations are quite different, as the former imposes an energy jump between the two phases while the latter implements a smooth transition between the phases. It is worth mentioning that although the MC has been extensively applied in the literature, the GC have not received the attention it deserves (for a recent relevant study see Ref.~\cite{Montana-2019}). In any case, the comparison of their predictions may offer useful physical insight. 

Rotating neutron stars with exotic degrees of freedom in their cores have been already studied in limited papers by Banik {\it et al.}~\cite{Banik-2004}, Bhattacharyya {\it et al.}~\cite{Bhattacharyya-2005},   Haensel {\it et al.}~\cite{Haensel-2016}, and Bozzola {\it et al.}~\cite{Bozzola-2019}. In the present study, in addition to the mentioned similar studies, the effects of rapid rotation on the twin stars scenario have been explored. The reason for this kind of study is twofold: on one hand we intend to examine to what extent the high rotation can differentiate the representation of the two stability branches of a static neutron star, and secondly to consider the possible cases where the existence of the twin stars may have arisen as a result of the high rotational frequency, where in the corresponding static case such a scenario is not foreseen. According to our knowledge this peculiar case has not been explored in details in the relevant bibliography, at least up to now. There is only an interesting mention in Ref.~\cite{Banik-2004}. Nonetheless, future study is in progress, in order to further clarify the effects of rotation on twin stars problem.

Finally, we employ the constraints on mass and radius  of the recently observed  compact object within the supernova remnant HESS J1731-347~\cite{Doroshenko-2022}. The corresponding estimations  are $M=0.77_{-0.17}^{+0.20}\ M_{\odot}$ and $R=10.4_{-0.78}^{+0.86} {\rm km}$, respectively, while the observations were carried out with the help  of  modelling of the X-ray spectrum and a robust distance estimate from Gaia observations~\cite{Doroshenko-2022}. According to the authors' guess, the above estimates imply that this object is either the lightest neutron star known, or a star with a more exotic equation of state. In any case, it is worth considering to what extent this compact object is compatible with the hybrid model and thus with the twin stars theory.

The paper is organized as follows. In Sec.~\ref{sec:HQPT} we present the basic formalism of the hadron to quark phase transition, while in Sec.~\ref{sec:tidal} we provide the tidal deformability. Sec.~\ref{sec:Results} is dedicated to the presentation and discussion of the results of the present study and, finally, the concluding remarks are given in Sec.~\ref{sec:Remarks}.

\section{Hadronic to quark matter phase transition} \label{sec:HQPT}

\subsection{Hybrid EOS and 1st-Order Phase Transition}
The theoretical framework throughout this work is based on the two prescriptions for matching the low-intermediate density hadronic EOS with the one describing the free quark matter, the Maxwell and the Gibbs  constructions (for an extensive and insight analysis see the recent review~\cite{Baym-2018}). In particular, while both methods have a phenomenological origin, they mimic, in a way, the theoretical approach of phase transition. It is important to note that both these constructions, do not take into account finite-size effects in the theory, so there are no Surface or Coulomb terms contributions present. In this sense, they should be considered as a useful mathematical tool rather than an exact description of the phase transition from hadronic to free quark matter. Furthermore, although the first method (MC) appears to be more applicable, the recent observations of gravitational waves related to the binary neutron star merger GW170817 event, along with the flexibility of the second method (GC), have made the latter case a more compelling candidate~\cite{Montana-2019}. Nevertheless, both methods have been used in the literature, leading under certain circumstances to the appearance of twin stars.

We note here that as the quark phase has a strong dependence on the speed of sound, for both configurations, in order to achieve the observed neutron star masses, at the selected transition densities, the speed of sound is set equal to the speed of light, having as a result the maximally stiff EOS.

\subsubsection{Maxwell construction}
The first case is the well-known MC case, which exhibits a sharp transition at the boundary and makes it hard for charged clusters of quarks to form in the hadronic matter. This particular construction is the favored one, in the case where the surface tension $\sigma_s$ in the hadron-quark crossover, is higher than the critical value of $\mathrm{\approx 40 \;MeV\;fm^{-3}}$ and less than the maximum allowed one of $\mathrm{\approx 100 \; MeV\;fm^{-3}}$ according to QCD lattice gauge simulations, with its exact value highly uncertain~\cite{Mariani-2017}. Of course, as it is expected, an abrupt phase change of this kind, yields a discontinuity in at least one physical quantity, with the most obvious one, the energy density, claiming the form~\cite{Alford-2013,Christian-2018,Montana-2019}
\begin{equation}
  \mathcal{E}(P) = \begin{cases} 
      \mathcal{E}_{\rm HADRON}(P), & P\leq P_{\rm tr} \\
      \mathcal{E}(P_{{\rm tr}}) + \Delta \mathcal{E} + c_s^{-2}(P-P_{{\rm tr}}), & P > P_{{\rm tr}}.
   \end{cases}
   \label{MC-1}
\end{equation}
In the above formula $\mathcal{E}(P)$ denotes the energy density, $P$ the pressure, $c_{\rm s}=\sqrt{\partial P / \partial \mathcal{E}}$ the speed of sound in units of speed of light, and $\Delta \mathcal{E}$ the magnitude of the energy density jump at the transition point. 
During the quark phase the numerical value we assign for $c_{\rm s}$ is equal to $c_s=1$, the maximum allowed value that is consistent with causality. That way, we also ensure the stiffest EOS case and the greatest possible maximum mass in the resulting M-R diagram. Moreover, $P_{\rm tr}$ expresses the pressure that corresponds to the baryon density at phase transition point, $n_{\rm tr}$. It should be clear that the first line in Eq.~\eqref{MC-1} refers to the hadronic phase while the second one refers to the quark phase, seen as the 1st-order Taylor expansion of the energy density around the transition pressure plus the $\Delta \mathcal{E}$ term. This is the so-called constant speed of sound parametrization (CSS)~\cite{Alford-2013,Montana-2019,Margaritis-2020,Kanakis-2020,Koliogiannis-2021f}. The described process requires not just the pressure but also the rest of the intensive thermodynamic quantities, meaning temperature T (set equal to 0) and baryonic chemical potential $\mu_{\rm B}$, of both phases to have the same value at the phase transition. The same is not true for the electric chemical potential $\mu_{\rm Q}$, as there is a jump at the interface between the two phases~\cite{Bhattacharyya-2010}. Local charge neutrality conditions for both regions separately, must be imposed, in order to ensure $\beta$-equilibrium state.

\subsubsection{Gibbs construction}
In the case of a non-sharp HQPT, meaning very low values of the surface tension $\sigma_s$, a finite region to embody the transition is implied. The mixed phase of this transition is composed, as its name suggests, by intermittent domains of pure hadronic and quark phases~\cite{Bhattacharyya-2010,Montana-2019}. The Gibbs phase transition rule regarding the equality of the pressure of the two components (Hadron - Intermediate, Intermediate - Quark), is established here. Constraints requiring global charge neutrality and baryon conservation number are imposed throughout the whole process, meaning that both baryon and electric chemical potentials, $\mu_{\rm B}$ and $\mu_{\rm Q}$, are required to have the same value across the phase boundaries. This is quite important, as now both phases can be oppositely charged, as long as the mixed phase remains neutrally charged. In this case it is only logical that nuclear matter is positively charged, assuming equal number of protons and neutrons (minimizing nuclear asymmetry energy), while to compensate, quark matter on the other hand has to be negatively charged. Contrary to the MC case, where the pressure remains constant in the transition interval, in the GC case, the pressure increases with increasing baryon density, while also no discontinuities in the energy density appear, giving rise to the profile~\cite{Montana-2019}
\begin{equation}
  \mathcal{E}(P) = \begin{cases} 
      \mathcal{E}_{\rm HADRON}(P), & P\leq P_{\rm tr}, \\
      A_{\rm m}\left({P}/{K_{\rm m}}\right)^{1/\Gamma_{\rm m}}+
      \gamma_{\rm m} P, & P_{\rm tr} \leq P \leq P_{\rm CSS}, \\
      \mathcal{E}(P_{{\rm CSS}}) + c_s^{-2}(P-P_{{\rm CSS}}), & P \geq P_{\rm CSS},
   \end{cases}
   \label{GC-1}
\end{equation}
where $A_{\rm m} = 1 + \alpha_{\rm m}$, $\gamma_{\rm m} = \left(\Gamma_{\rm m} -1\right)^{-1}$, and $c_{\rm s}=1$ (maximally stiff EOS). The energy density is denoted by $\mathcal{E}(P)$, the pressure by $P$, the speed of sound by $c_{\rm s}$, while $a_m$, $K_m$ and the polytropic index $\Gamma_m$ are constants, with the former two evaluated, by requiring continuity of $P$ and $\mathcal{E}$ at the transition points. As has already been explained, an energy density jump is not present here, opposed to the MC, but the increase in the quantity is assigned during the intermediate part of the formula. Subscript $``{\rm \rm css}"$ denotes the corresponding quantity at the start of the quark phase. That being said, it should be understood that discontinuities in the derivatives of the pressure with regards to the energy density are still present here in the points where the mixed phase begins and ends. This is clearly reflected in the value of the speed of sound $c_s$ between the three phases.

The relation between the pressure and baryon density for the intermediate section in Eq.~\eqref{GC-1} is given through the well-known, simple polytropic formula $P(n)=K_m n^{\Gamma_m}$. While the values of $a_m$ and $K_m$ depend on the transition point, the polytropic index $\Gamma_m$ in the GC is taken to be constant throughout the whole work and equal to 1.03. A value so close to unity is justified by taking into account the uncertainty related to how soft the EOS of the mixed phase is and also differentiating it as much as possible from the respective value of the MC, where according to the relevant discussion above, we have $\Gamma_m=0$ (see also Ref.~\cite{Montana-2019}).

\subsection{Seidov Criterion}
In general, the phase transition from the hadronic to quark matter described in the previous section, is not sufficient by itself to predict the existence of a third family of compacts stars. To be more specific, this existence requires, in the corresponding mass-radius diagram, the appearance of an unstable region followed by a stable one (for an instructive discussion see Ref.~\cite{Bielich-2020}). The criterion for causing an unstable region by a first-order phase transition in neutron stars was firstly considered by Seidov~\cite{Seidov-1971}. His work was based on the work of Lighthill~\cite{Lighthill-1950}, by performing a linear perturbative expansion in the size of the quark matter core.

In particular, we consider an EOS and a phase transition at the pressure $P_{\rm tr}$ which corresponds to the energy density $\mathcal{E}_1\equiv \mathcal{E}_{\rm tr}$ with  a simultaneous jump to the energy density $\mathcal{E}_2\equiv \mathcal{E}(P_{{\rm tr}}) + \Delta \mathcal{E}$. In general, a stable sequence where the mass increases with the central pressure will become an unstable one as the mass decreases with the central pressure. This case takes place only if the following inequality holds~\cite{Bielich-2020}
\[ 3 P_{\rm tr }+3\mathcal{E}_1-2\mathcal{E}_2<0.  \]
Thus, we finally find the minimum jump in the energy density that is required for the appearance of an unstable configuration and which is given by the relation
\begin{equation}
\Delta \mathcal{E}_{\rm cr}=\frac{1}{2}\mathcal{E}_{\rm tr}+\frac{3}{2}P_{\rm tr}.
\label{eq:seidov_limit}
\end{equation}
In order to have a third family of compact objects appear in the M-R diagram, the aforementioned instability has to be satisfied. The theory dictates that as long as the mass M is an increasing function of the central pressure $P_c$, the star will remain stable. As the central pressure increases, it reaches a certain point where its value becomes equal to the transition pressure $P_{\rm tr}$, leading to the formation of the quark matter core. Depending on whether the Seidov criterion for the energy density jump is surpassed or not, there are four possible outcomes, with only two of them producing a new type of compact object~\cite{Alford-2013,Bielich-2020}. It should be noted here that the Seidov criterion is meaningful only in the presence of a sharp discontinuity in the energy density profile, which appears in the MC method as opposed to the GC method.

In order to be able to compare the GC with the MC, where an energy jump appears in the form of $\Delta \mathcal{E}_{\rm cr}$, we define the corresponding energy increase in the GC as the quantity
\begin{equation}
    \Delta \mathcal{E}_{\rm G}= \frac{3}{2} \left(\frac{1}{2}\mathcal{E}_{\rm HADRON}(P_{\rm tr})+\frac{3}{2}P_{\rm tr}\right),
\end{equation}
with $\mathcal{E}_{\rm HADRON}(P_{\rm tr})$ and $P_{\rm tr}$ representing the respective values at the transition from the hadron phase to the mixed phase.

The two possible outcomes for producing a new type of compact objects are briefly  described below.  Case 1: If $\Delta\mathcal{E}$ is taken to be much higher than the Seidov limit ($\Delta \mathcal{E}>>\Delta \mathcal{E}_{\rm cr}$), then the instability appears immediately after the formation of the quark matter core. On the M-R diagram, the transition point takes the form of a cusp, meaning the sign of $dM/dR$ is flipped. This case does not result to the formation of a stable hybrid star (see case $\mathrm{n_{tr}}=0.5 \; \mathrm{fm^{-3}}$ in Figure \ref{fig:MR_maxwell_GRDF-DD2}). Case 2: The value of $\Delta \mathcal{E}$ should be higher that the value of Eq.~\eqref{eq:seidov_limit} ($\Delta \mathcal{E}>\Delta \mathcal{E}_{\rm cr}$), but not high enough to reach Case 1 scenario.  This case describes the appearance of a third family of compact objects.  The cusp is also present here during the transition point, leading to an unstable disconnected area in the graph, which after a while flips around again, signalling the appearance of a stable hybrid star with a quark matter core.  

In the present work we study also  two  cases where the Seidov criterion is violated (for more details see chapter 9 of Ref.~\cite{Bielich-2020} and Fig.3 of Ref.~\cite{Alford-2013}). They both include the hybrid branch connected to the nuclear star branch that may lead to the appearance of a third family of compact objects. These two cases are described below with their corresponding numbering as a continuation of the two previous cases. Case 3: The energy density jump $\Delta \mathcal{E}$ is not high enough to cause an instability at the moment ($\Delta \mathcal{E}<\Delta \mathcal{E}_{\rm cr}$), meaning that the quark matter is described by a curve on the map that is connected to the hadronic matter branch. The curve continues up to a point, after which an unstable region appears leading again to a stable hybrid branch, just like Case 2 scenario, albeit shorter in length (see Figure~\ref{fig:MR_nsl_5_GRDF-DD2_MC} for $\mathrm{n_{tr}}=0.38 \; \mathrm{fm^{-3}}$). Case 4:  The hybrid star branch connected to the nuclear branch is also present here with its length to be inversely proportional to the ratio of $\Delta \mathcal{E}/\mathcal{E}_{\rm tr}$. It is shown that higher values of the above quantity, lead to a decrease in the size of the hybrid branch that spans the phase diagram. Immediately after a maximum mass value is reached, a sudden discontinuity appears that continues forth without the appearance of a third family of compact objects~\cite{Alford-2013}.

\section{Tidal deformability} \label{sec:tidal}
A very important source for the gravitational wave detectors are the gravitational waves from the late phase of the inspiral of a binary neutron star system, before the merger~\cite{Postnikov-2010,Flanagan-08,Hinderer-08}. This kind of source leads to the measurement of various properties of neutron stars. In the inspiral phase the tidal effects can be detected~\cite{Flanagan-08}.

The $k_2$ parameter, also known as tidal Love numbeer, depends on the equation of state and describes the response of a neutron star to the tidal field $E_{ij}$~\cite{Flanagan-08}. The exact relation is given below
\begin{equation}
Q_{ij}=-\frac{2}{3}k_2\frac{R^5}{G}E_{ij}\equiv- \lambda E_{ij},
\label{Love-1}
\end{equation}
where $R$ is the neutron star radius and $\lambda=2R^5k_2/3G$ is the tidal deformability. The tidal Love number $k_2$ is given by \cite{Flanagan-08,Hinderer-08}
\begin{eqnarray}
k_2&=&\frac{8\beta^5}{5}\left(1-2\beta\right)^2\left[2-y_R+(y_R-1)2\beta \right]\nonumber\\
& \times&
\left[\frac{}{} 2\beta \left(6  -3y_R+3\beta (5y_R-8)\right) \right. \nonumber \\
&+& 4\beta^3 \left.  \left(13-11y_R+\beta(3y_R-2)+2\beta^2(1+y_R)\right)\frac{}{} \right.\nonumber \\
&+& \left. 3\left(1-2\beta \right)^2\left[2-y_R+2\beta(y_R-1)\right] {\rm ln}\left(1-2\beta\right)\right]^{-1},
\label{k2-def}
\end{eqnarray}
where $\beta=GM/Rc^2$ is the compactness of a neutron star. The paramter $y_R$ is determined by solving numerically the following differential equation
\begin{equation}
r\frac{dy(r)}{dr}+y^2(r)+y(r)F(r)+r^2Q(r)=0, 
\label{D-y-1}
\end{equation}
with the initial condition $ y(0)=2$~\cite{Hinderer-10}. $F(r)$ and $Q(r)$ are functionals of the energy density ${\cal E}(r)$, pressure $P(r)$, and mass $M(r)$ defined as~\cite{Postnikov-2010}
\begin{equation}
F(r)=\left[ 1- \frac{4\pi r^2 G}{c^4}\left({\cal E} (r)-P(r) \right)\right]\left(1-\frac{2M(r)G}{rc^2}  \right)^{-1},
\label{Fr-1}
\end{equation}
and
\begin{eqnarray}
r^2Q(r)&=&\frac{4\pi r^2 G}{c^4} \left[5{\cal E} (r)+9P(r)+\frac{{\cal E} (r)+P(r)}{\partial P(r)/\partial{\cal E} (r)}\right]
\nonumber\\
&\times&
\left(1-\frac{2M(r)G}{rc^2}  \right)^{-1}- 6\left(1-\frac{2M(r)G}{rc^2}  \right)^{-1} \nonumber \\
&-&\frac{4M^2(r)G^2}{r^2c^4}\left(1+\frac{4\pi r^3 P(r)}{M(r)c^2}   \right)^2\left(1-\frac{2M(r)G}{rc^2}  \right)^{-2}.
\label{Qr-1}
\end{eqnarray}
Eq.~(\ref{D-y-1}) must be solved numerically and self consistently with the Tolman - Oppenheimer - Volkoff (TOV) equations under the following boundary conditions: $y(0)=2$, $P(0)=P_c$ ($P_{c}$ denotes the central pressure), and $M(0)=0$~\cite{Postnikov-2010,Hinderer-08}. From the numerical solution of TOV equations, the mass $M$ and radius $R$ of the neutron star can be computed, while the corresponding solution of the differential Eq.~(\ref{D-y-1}) provides the value of $y_R=y(R)$. The last parameter along with the quantity $\beta$ are the  basic ingredients  of the tidal Love number $k_2$.

The chirp mass {\it $\mathcal{M}_c$} of a binary neutron stars system is a well measured quantity by the gravitational wave detectors~\cite{Abbott-gw170817}. Its relation is given below
\begin{equation}
\mathcal{M}_c=\frac{(m_1m_2)^{3/5}}{(m_1+m_2)^{1/5}}=m_1\frac{q^{3/5}}{(1+q)^{1/5}},
\label{chirpmass}
\end{equation}
where $m_1$ is the mass of the heavier component star and $m_2$ is the lighter's one. Hence, the binary mass ratio $q=m_2/m_1$ lies within the range $0 < q\leq1$.

Additionally, another quantity that is well measured is the effective tidal deformability $\tilde{\Lambda}$ which is given by~\cite{Abbott-gw170817}
\begin{equation}
\tilde{\Lambda}=\frac{16}{13}\frac{(12q+1)\Lambda_1+(12+q)q^4\Lambda_2}{(1+q)^5},
\label{L-tild-1}
\end{equation}
where $\Lambda_i$ is the dimensionless tidal deformability~\cite{Abbott-gw170817}
\begin{equation}
\Lambda_i=\frac{2}{3}k_2\left(\frac{R_i c^2}{M_i G}  \right)^5\equiv\frac{2}{3}k_2 \beta_i^{-5}  , \quad i=1,2.
\label{Lamb-1}
\end{equation}
The effective tidal deformability $\tilde{\Lambda}$ is one of the main quantities that can be well measured by the detection of the corresponding gravitation waves.


\begin{figure*}
    \centering
    \includegraphics[width=\columnwidth]{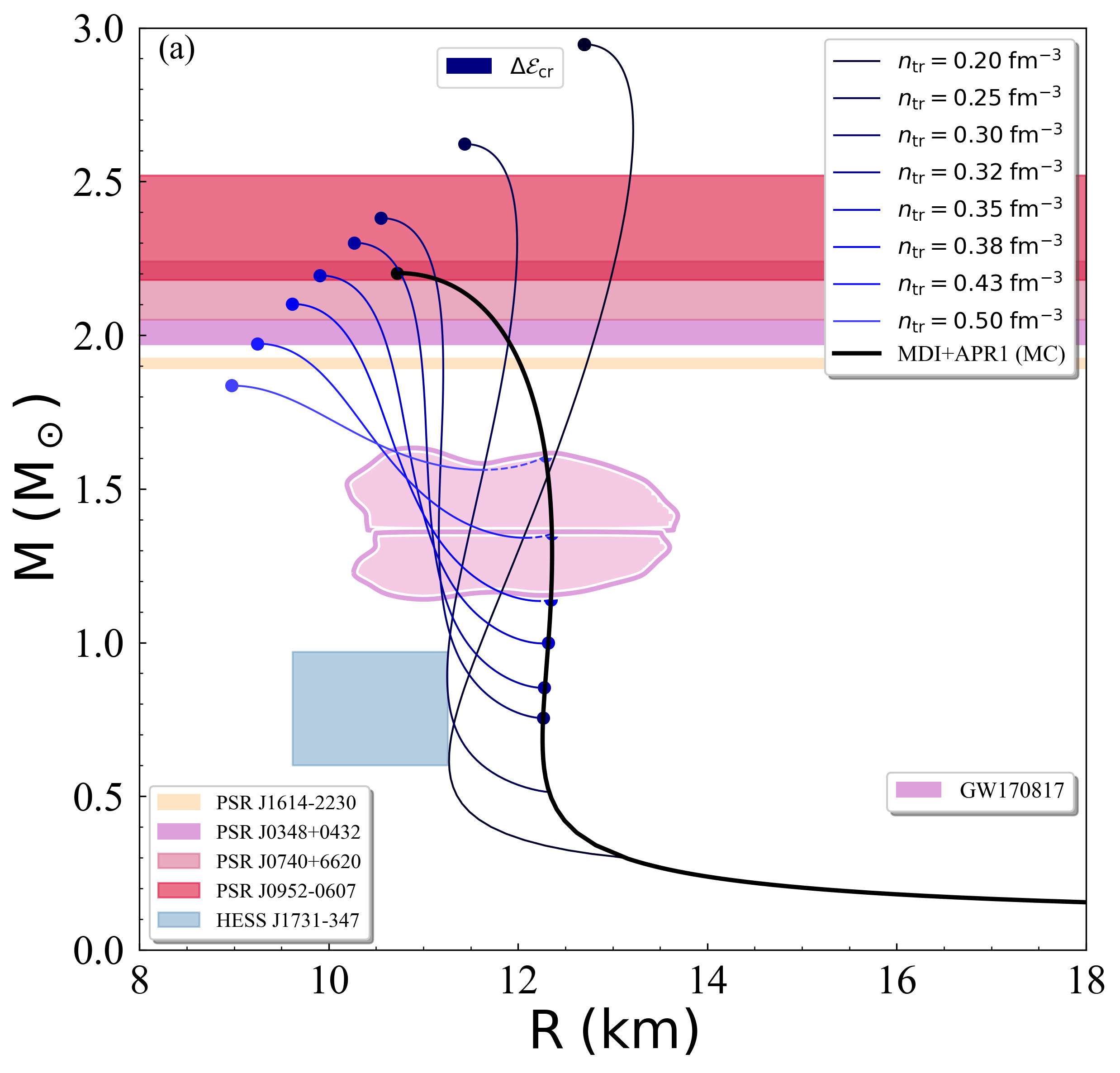}~
    \includegraphics[width=\columnwidth]{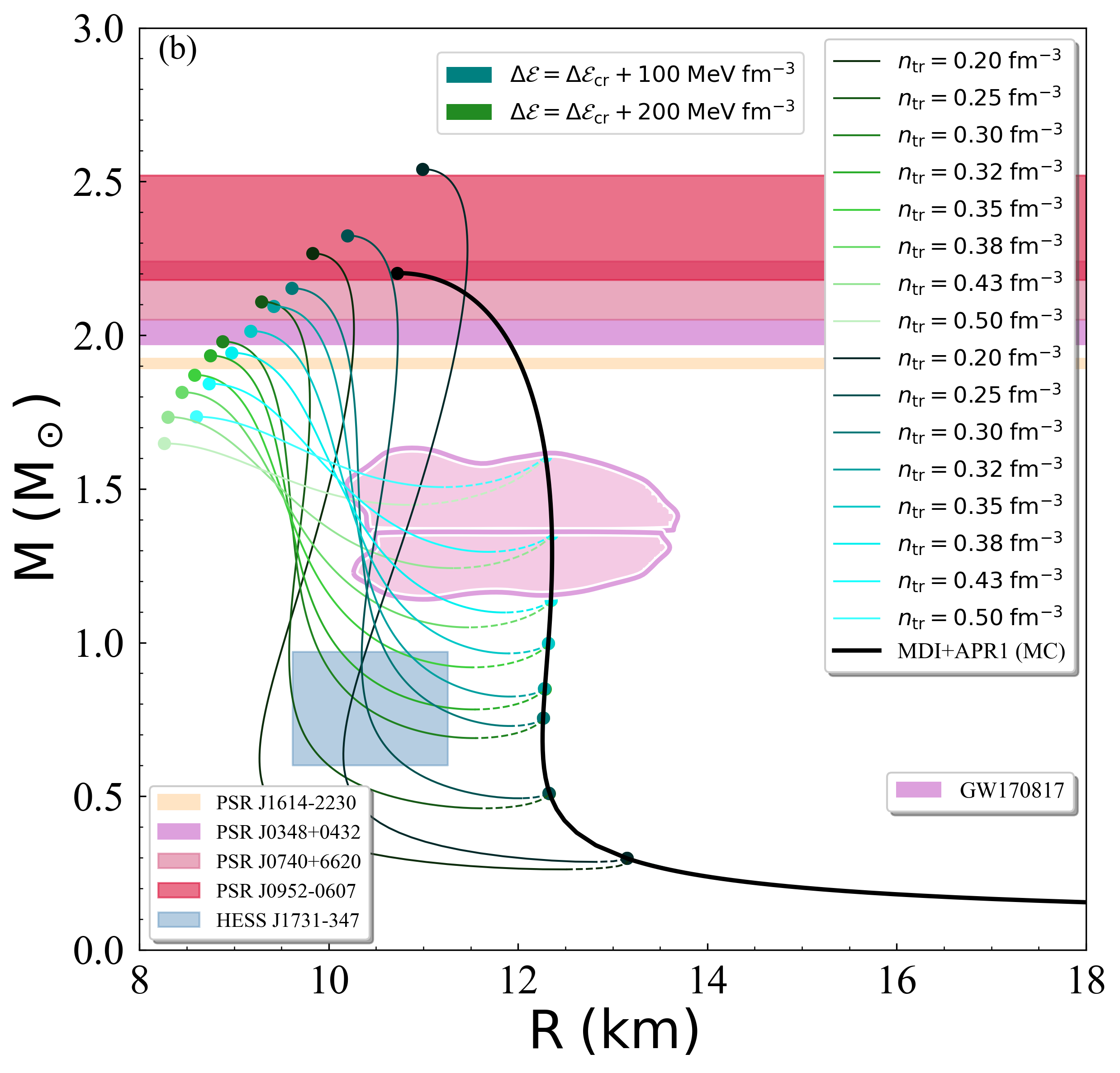}
      \caption{Mass vs Radius diagram for the MDI+APR1 EOS under MC and for (a) $\Delta\mathcal{E}_{\rm cr}$, and (b) $\mathrm{\Delta\mathcal{E}=\Delta\mathcal{E}_{\rm cr}+100\;MeV\;fm^{-3}}$ (blue curves) and $\mathrm{\Delta\mathcal{E}=\Delta\mathcal{E}_{\rm cr}+200\;MeV\;fm^{-3}}$ (green curves). The black curve indicates the original EOS. The shaded regions from bottom to top represent the HESS J1731-347 remnant~\cite{Doroshenko-2022}, GW170817 event~\cite{Abbott-gw170817}, PSR J1614–2230~\cite{Arzoumanian-2018}, PSR J0348+0432~\cite{Antoniadis-2013}, PSR J0740+6620~\cite{Cromartie-2020}, and PSR J0952-0607~\cite{Romani-2022} pulsar observations for possible maximum mass.}
    \label{fig:MR_maxwell_MDI}
\end{figure*}

\begin{figure*}
    \centering
    \includegraphics[width=\columnwidth]{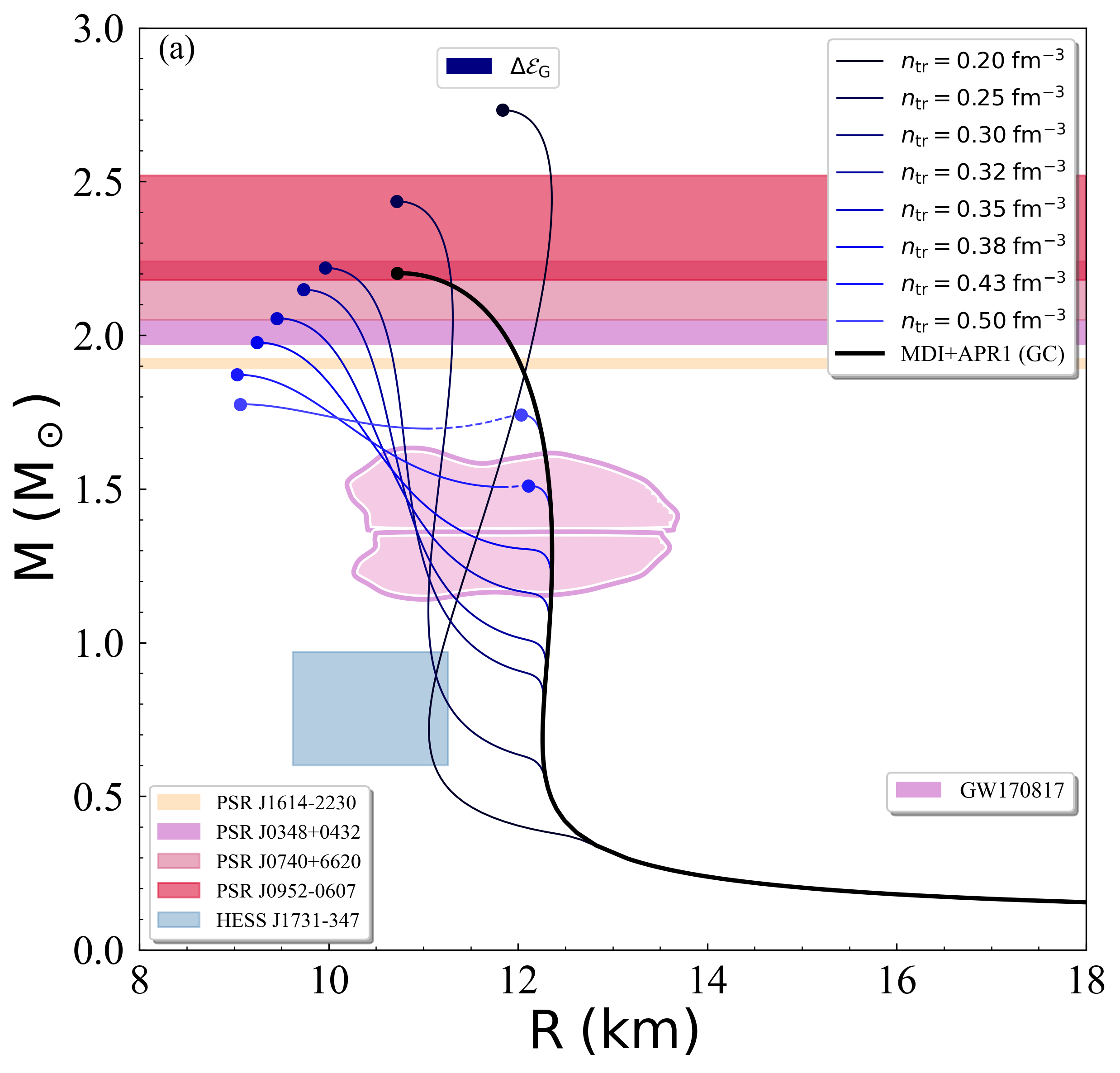}~
    \includegraphics[width=\columnwidth]{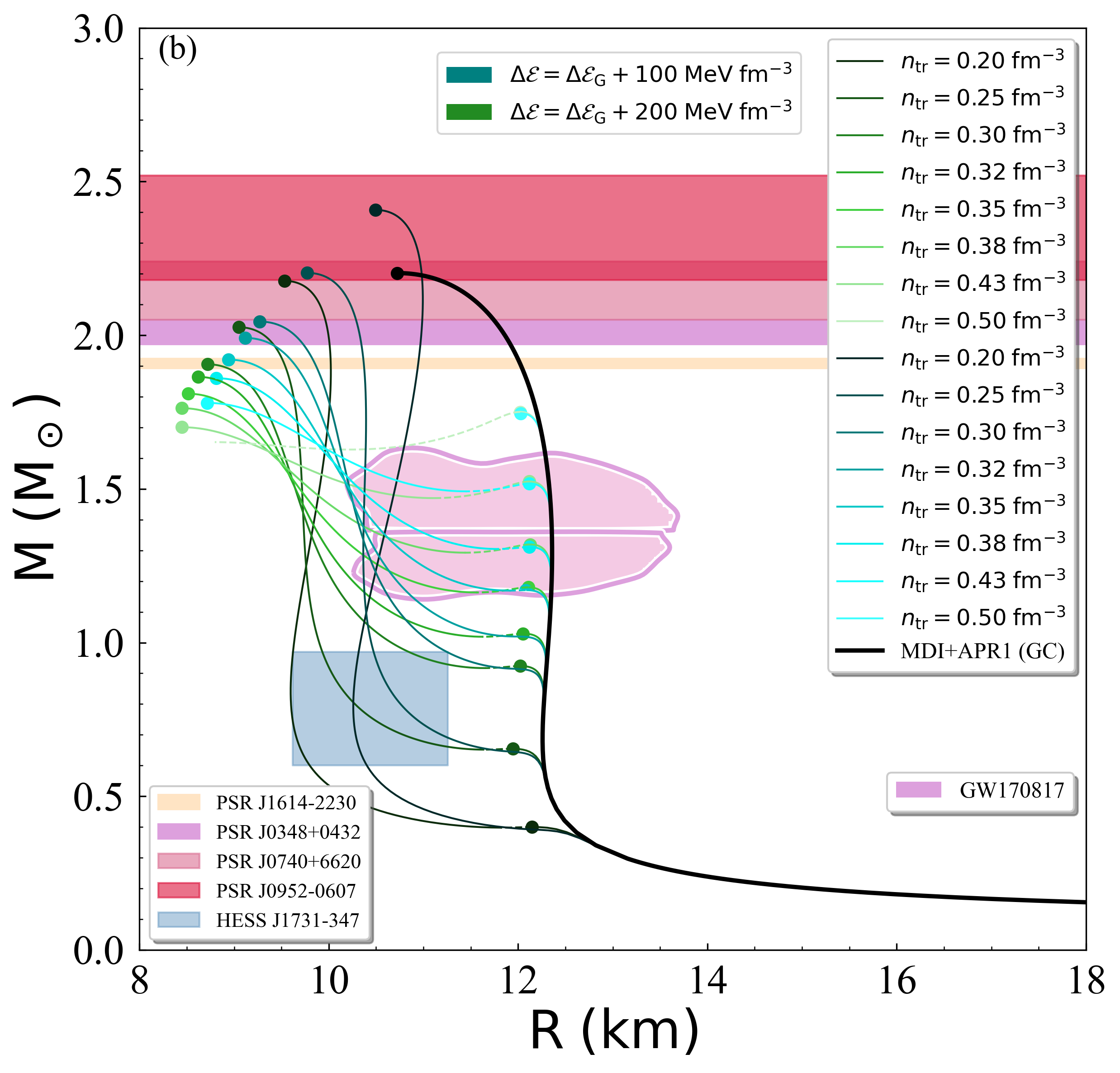}
      \caption{Mass vs Radius diagram for the MDI+APR1 EOS under GC and for (a) $\Delta\mathcal{E}_{\rm G}$, and (b) $\mathrm{\Delta\mathcal{E}=\Delta\mathcal{E}_{\rm G}+100\;MeV\;fm^{-3}}$ (blue curves) and $\mathrm{\Delta\mathcal{E}=\Delta\mathcal{E}_{\rm G}+200\;MeV\;fm^{-3}}$ (green curves). The black curve indicates the original EOS. The shaded regions from bottom to top represent the HESS J1731-347 remnant~\cite{Doroshenko-2022}, GW170817 event~\cite{Abbott-gw170817}, PSR J1614–2230~\cite{Arzoumanian-2018}, PSR J0348+0432~\cite{Antoniadis-2013}, PSR J0740+6620~\cite{Cromartie-2020}, and PSR J0952-0607~\cite{Romani-2022} pulsar observations for possible maximum mass.} 
    \label{fig:MR_gibbs_MDI}
\end{figure*}

\section{Results} \label{sec:Results}

\subsection{Mass vs Radius diagram}

\begin{figure*}
    \centering
    \includegraphics[width=\columnwidth]{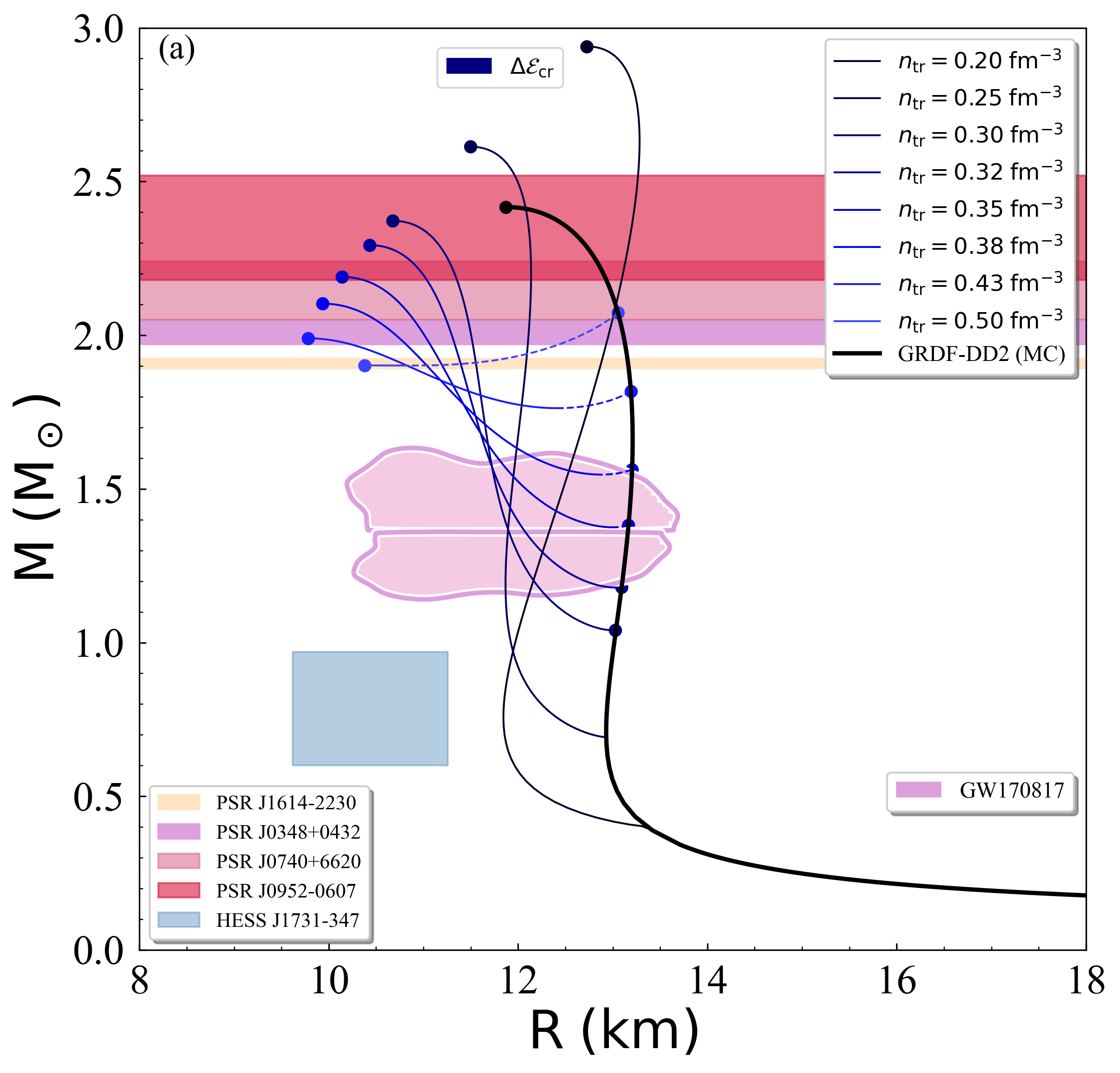}~
    \includegraphics[width=\columnwidth]{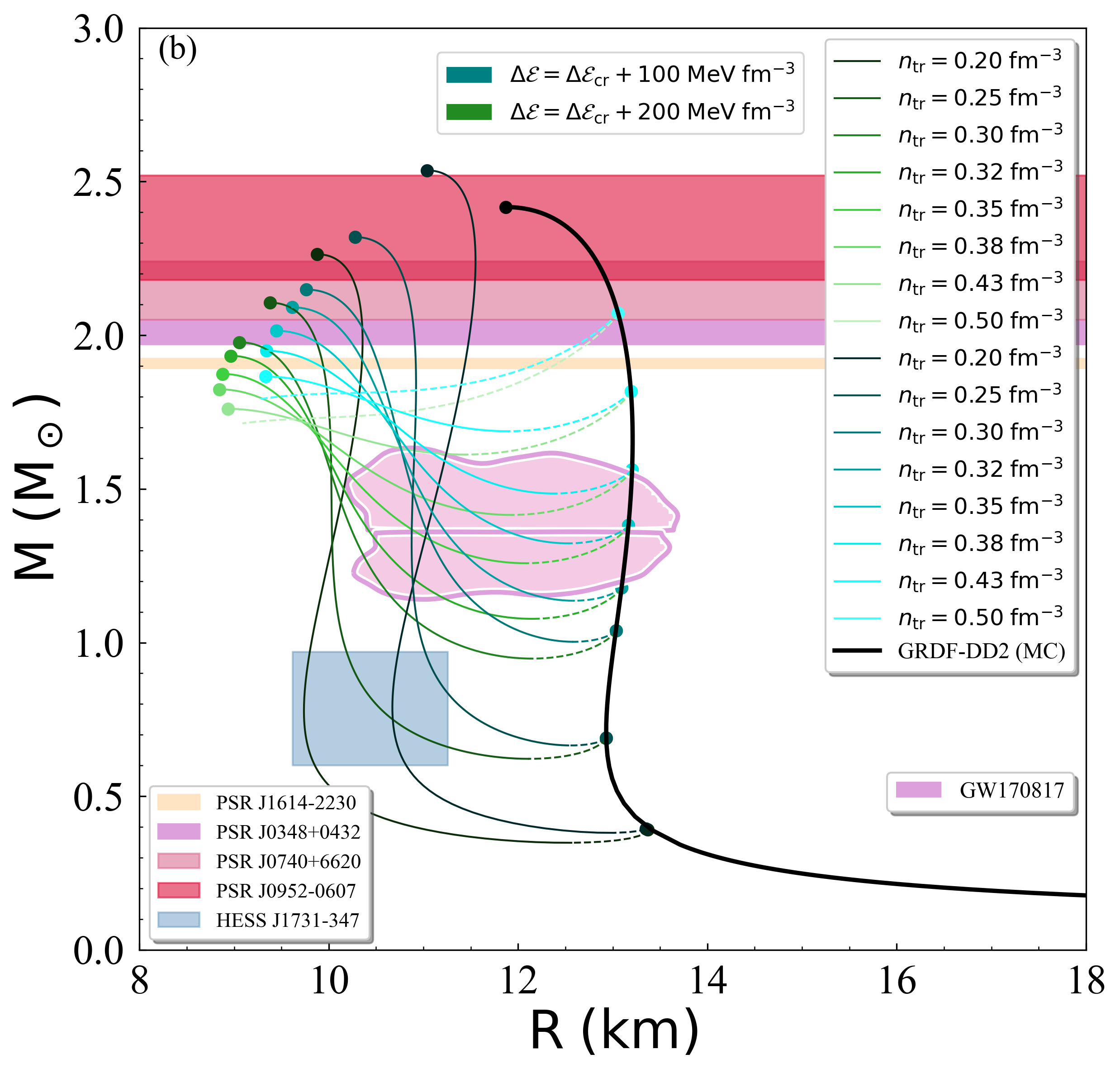}
      \caption{Mass vs Radius diagram for the GRDF-DD2 EOS under MC and for (a) $\Delta\mathcal{E}_{\rm cr}$, and (b) $\mathrm{\Delta\mathcal{E}=\Delta\mathcal{E}_{\rm cr}+100\;MeV\;fm^{-3}}$ (blue curves) and $\mathrm{\Delta\mathcal{E}=\Delta\mathcal{E}_{\rm cr}+200\;MeV\;fm^{-3}}$ (green curves). The black curve indicates the original EOS. The shaded regions from bottom to top represent the HESS J1731-347 remnant~\cite{Doroshenko-2022}, GW170817 event~\cite{Abbott-gw170817}, PSR J1614–2230~\cite{Arzoumanian-2018}, PSR J0348+0432~\cite{Antoniadis-2013}, PSR J0740+6620~\cite{Cromartie-2020}, and PSR J0952-0607~\cite{Romani-2022} pulsar observations for possible maximum mass.}
    \label{fig:MR_maxwell_GRDF-DD2}
\end{figure*}

\begin{figure*}
    \centering
    \includegraphics[width=\columnwidth]{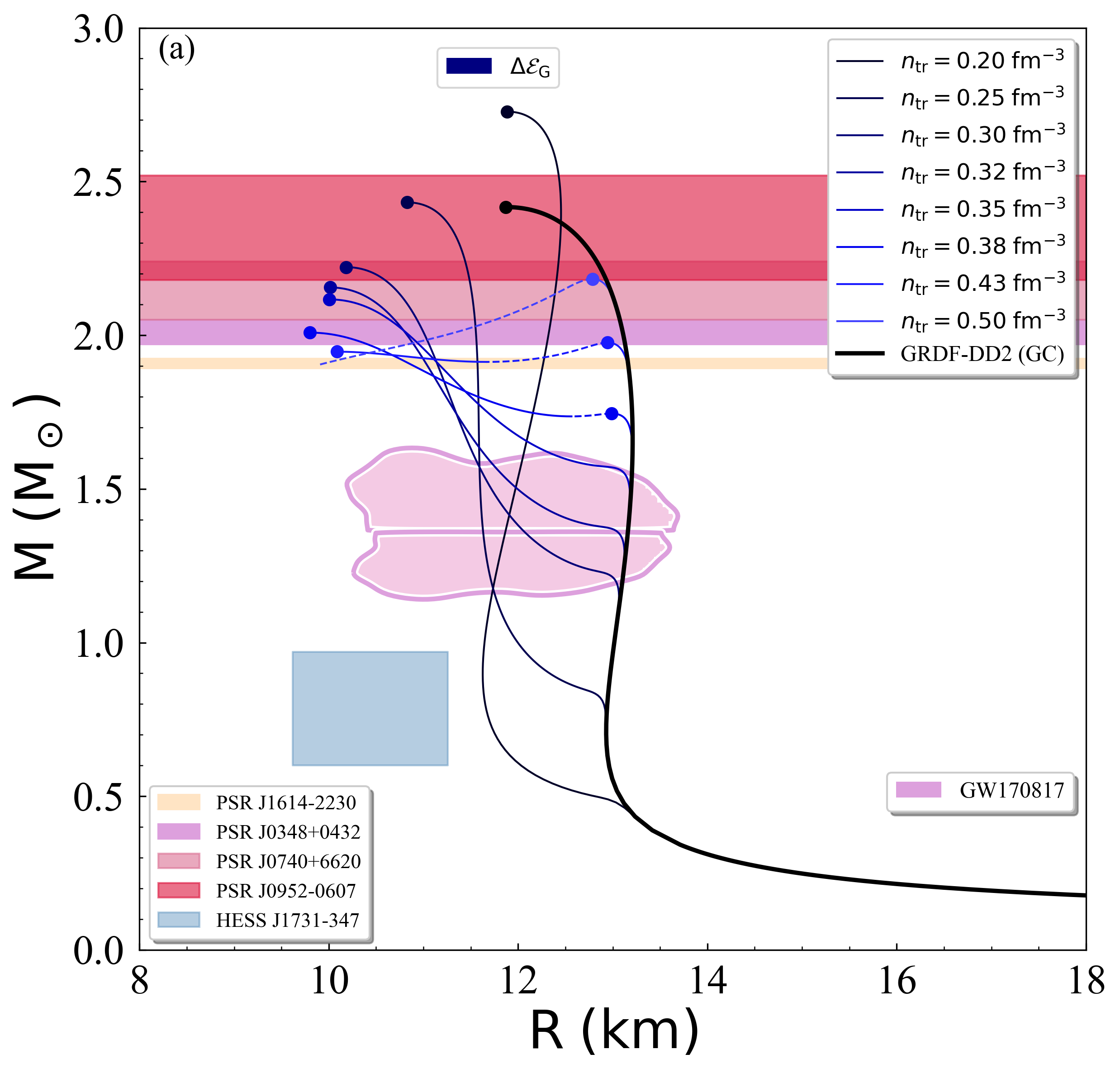}~
    \includegraphics[width=\columnwidth]{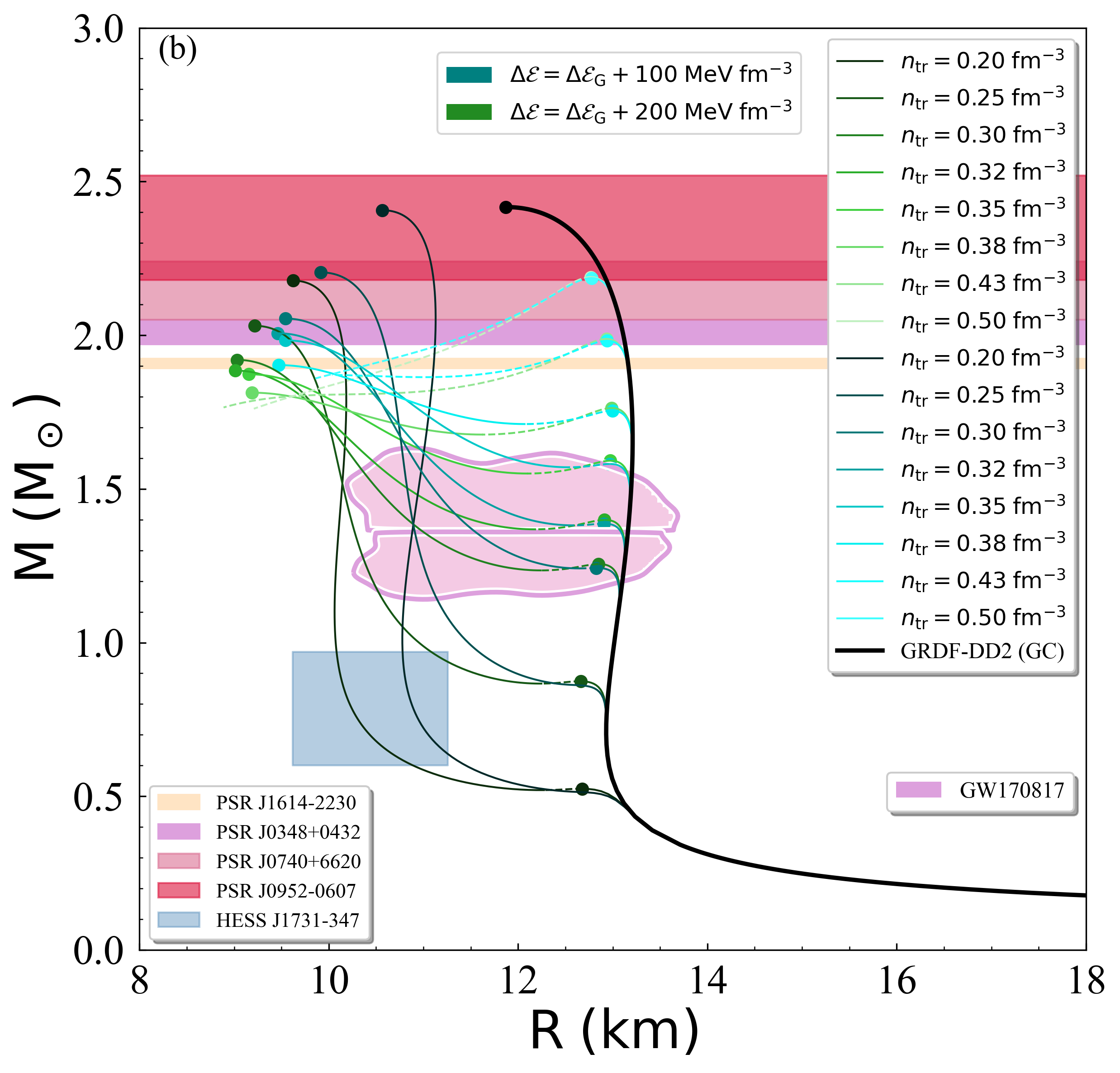}
      \caption{Mass vs Radius diagram for the GRDF-DD2 EOS under GC and for (a) $\Delta\mathcal{E}_{\rm G}$, and (b) $\mathrm{\Delta\mathcal{E}=\Delta\mathcal{E}_{\rm G}+100\;MeV\;fm^{-3}}$ (blue curves) and $\mathrm{\Delta\mathcal{E}=\Delta\mathcal{E}_{\rm G}+200\;MeV\;fm^{-3}}$ (green curves). The black curve indicates the original EOS. The shaded regions from bottom to top represent the HESS J1731-347 remnant~\cite{Doroshenko-2022}, GW170817 event~\cite{Abbott-gw170817}, PSR J1614–2230~\cite{Arzoumanian-2018}, PSR J0348+0432~\cite{Antoniadis-2013}, PSR J0740+6620~\cite{Cromartie-2020}, and PSR J0952-0607~\cite{Romani-2022} pulsar observations for possible maximum mass.}
    \label{fig:MR_gibbs_GRDF-DD2}
\end{figure*}

In our study, as we have already mentioned, we used two constructions, a) the MC, and b) the GC, and two different EOSs, a) the MDI+APR1 EOS~\cite{Koliogiannis-2021}, and b) the GRDF-DD2 EOS~\cite{Typel-2018}. Also, we focused on the following values of $\mathrm{n_{tr}}=[0.20,0.25,0.30,0.32,0.35,0.38,0.43,0.50] \; \mathrm{fm^{-3}}$. In Figures~\ref{fig:MR_maxwell_MDI}-\ref{fig:MR_gibbs_GRDF-DD2} we show the M-R diagrams for all cases. In particular, Figures~\ref{fig:MR_maxwell_MDI} and~\ref{fig:MR_gibbs_MDI} indicate the MDI+APR1 EOS with the MC and GC, respectively, while Figures~\ref{fig:MR_maxwell_GRDF-DD2} and~\ref{fig:MR_gibbs_GRDF-DD2} show the GRDF-DD2 EOS with the MC and GC, respectively. In each Figure, the left panel (a) corresponds to $\Delta \mathcal{E}$ ($\Delta \mathcal{E}_{\rm cr}$ for MC and $\Delta \mathcal{E}_{\rm G}$ for GC) and the right panel (b) corresponds to the cases $\mathrm{\Delta\mathcal{E}=\Delta \mathcal{E}_{cr}+[100,200]\; MeV\;fm^{-3}}$ for MC and $\mathrm{\Delta\mathcal{E}=\Delta \mathcal{E}_{G}+[100,200]\; MeV\;fm^{-3}}$ for GC respectively. A general remark is that the EOSs that fulfill the $\Delta \mathcal{E}_{\rm cr}$ ($\Delta \mathcal{E}_{\rm G}$) are stiffer than the cases with $\mathrm{\Delta\mathcal{E}=\Delta \mathcal{E}_{cr}+[100,200]\; MeV\;fm^{-3}}$ ($\mathrm{\Delta\mathcal{E}=\Delta \mathcal{E}_{G}+[100,200]\; MeV\;fm^{-3}}$). Especially, as we move to higher values of $\mathrm{\Delta\mathcal{E}}$, the EOSs become softer (smaller higher masses and smaller radii). The purple horizontal shaded regions correspond to observational data from pulsars, while the light purple contour shaded region indicates the observation of the GW170817 event~\cite{Abbott-gw170817}. The dashed part of curves indicates their unstable region. Another remark is that between the two constructions, MC and GC, the first one provides branches of EOSs in a lower mass region than the last one. Therefore, the MC is more informative in our case of study, since the GW170817 event contains low values of component masses. Also, as one can observe, as we move to higher values of $\mathrm{n_{tr}}$ the EOSs become softer.

Moreover, in Figures~\ref{fig:MR_maxwell_MDI}-\ref{fig:MR_gibbs_GRDF-DD2},  the constraints on mass and radius  of the recently  observed  remnant HESS J1731-347 have also been included~\cite{Doroshenko-2022}. Obviously, the prediction of the above constraints requires the use of a larger jump in the energy density for both constructions. In particular, we found that maybe this star is a hybrid star, belonging to the second stable branch. It is also remarkable that these predictions are compatible with the prediction of the maximum observed masses. It is worth  mentioning  that we do not claim that the  remnant HESS J1731-347 is definitely a hybrid star. However, according to the present study, this case is more favorable than to be a neutron star. Further similar observations are needed in order to clarify this issue.

\begin{figure*}
    \centering
    \includegraphics[width=\textwidth]{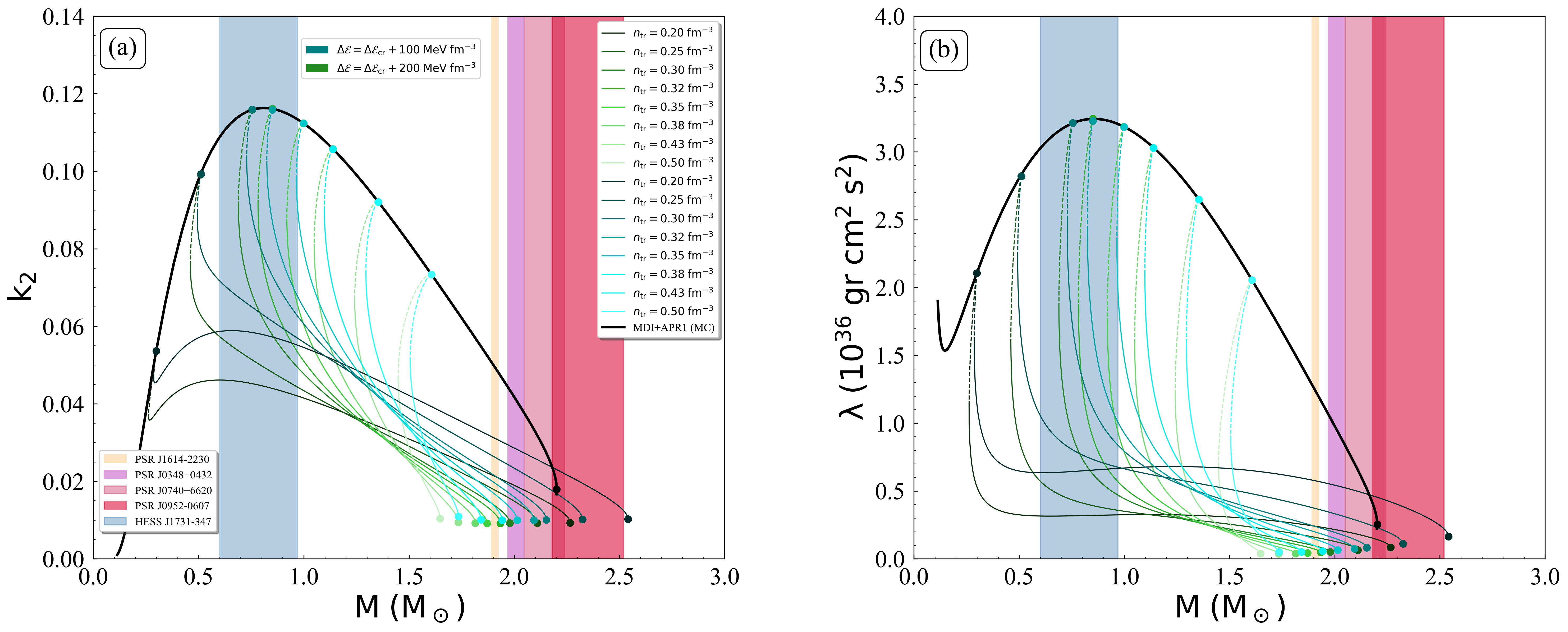}
    \caption{Tidal parameters (a) $\mathrm{k_2}$, and (b) $\lambda$ as a relation of the neutron star mass for the MDI+APR1 EOS under MC and for $\mathrm{\Delta\mathcal{E}=\Delta\mathcal{E}_{\rm cr}+100\;MeV\;fm^{-3}}$ (blue curves), and $\mathrm{\Delta\mathcal{E}=\Delta\mathcal{E}_{\rm cr}+200\;MeV\;fm^{-3}}$ (green curves). The black curve indicates the original EOS. The dashed part of curves indicates their unstable region. The shaded regions from left to right represent the HESS J1731-347 remnant~\cite{Doroshenko-2022}, PSR J1614–2230~\cite{Arzoumanian-2018}, PSR J0348+0432~\cite{Antoniadis-2013}, PSR J0740+6620~\cite{Cromartie-2020}, and PSR J0952-0607~\cite{Romani-2022} pulsar observations for possible maximum mass.}
    \label{fig:tidalparams_maxwell_MDI}
\end{figure*}

\begin{figure*}
    \centering
    \includegraphics[width=\columnwidth]{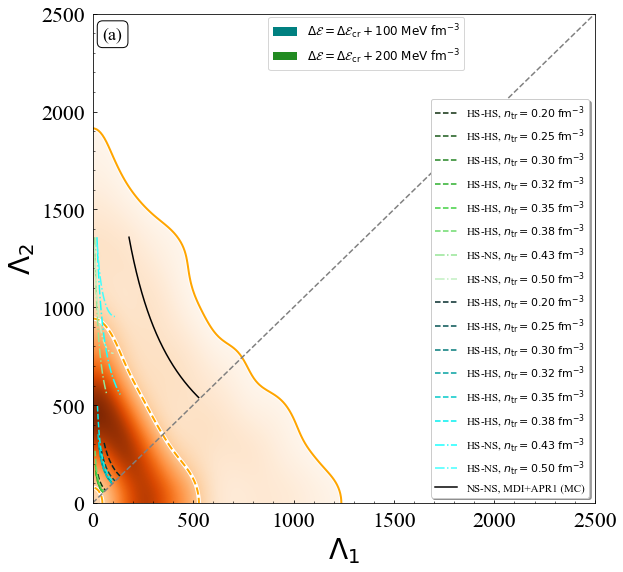}~
    \includegraphics[width=\columnwidth]{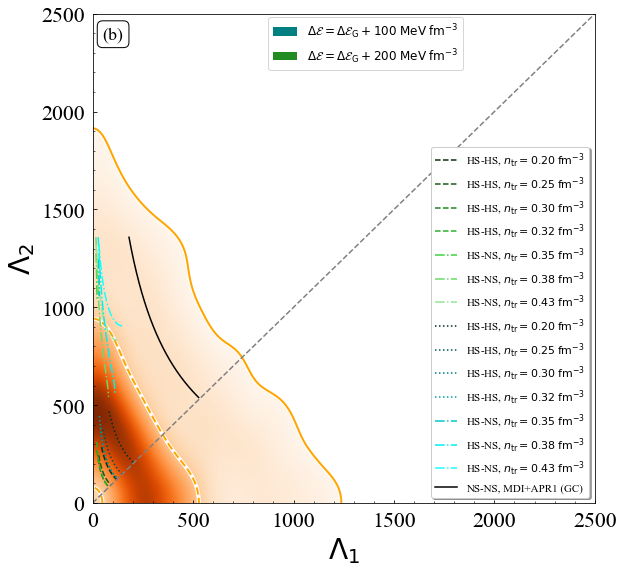}
    \caption{$\Lambda_1-\Lambda_2$ relation for the MDI+APR1 EOS and a) MC, and b) GC. The blue (green) curves correspond to the $\mathrm{\Delta\mathcal{E}=\Delta \mathcal{E}_{cr}+100\;MeV\;fm^{-3}}$ ($\mathrm{\Delta\mathcal{E}=\Delta \mathcal{E}_{cr}+200\;MeV\;fm^{-3}}$) for the MC, while for the GC correspond to the $\mathrm{\Delta\mathcal{E}=\Delta\mathcal{E}_{\rm G}+100\;MeV\;fm^{-3}}$ and $\mathrm{\Delta\mathcal{E}=\Delta\mathcal{E}_{\rm G}+200\;MeV\;fm^{-3}}$ respectively. The black curve indicates the original EOS. The shaded region shows the acceptance values derived by the GW170817 event~\cite{Abbott-gw170817}.}
    \label{fig:L1L2_MDI}
\end{figure*}

\begin{figure*}
    \centering
    \includegraphics[width=\columnwidth]{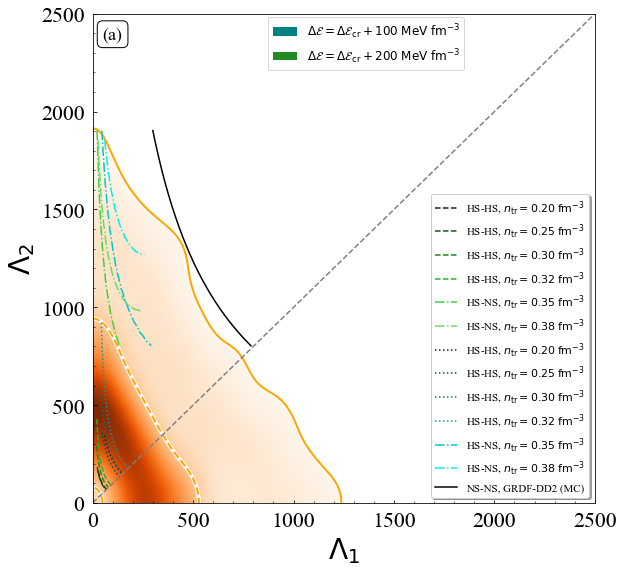}~
    \includegraphics[width=\columnwidth]{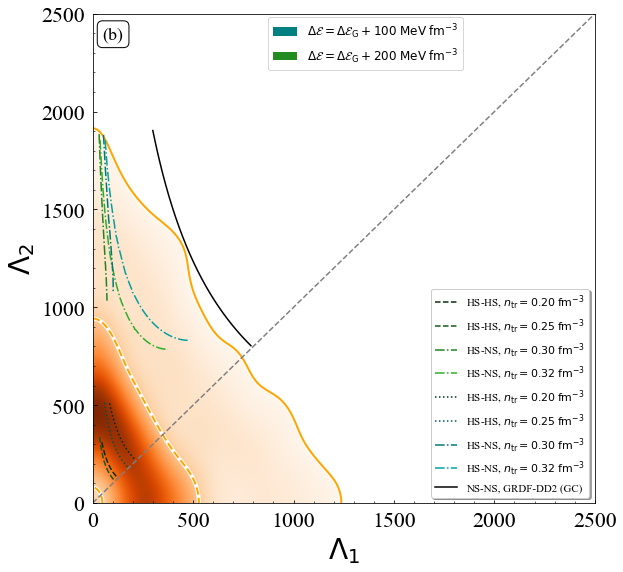}
    \caption{$\Lambda_1-\Lambda_2$ relation for the GRDF-DD2 EOS and a) MC, and b) GC. The blue (green) curves correspond to the $\mathrm{\Delta\mathcal{E}=\Delta \mathcal{E}_{cr}+100\;MeV\;fm^{-3}}$ ($\mathrm{\Delta\mathcal{E}=\Delta \mathcal{E}_{cr}+200\;MeV\;fm^{-3}}$) for the MC, while for the GC correspond to the $\mathrm{\Delta\mathcal{E}=\Delta\mathcal{E}_{\rm G}+100\;MeV\;fm^{-3}}$ and $\mathrm{\Delta\mathcal{E}=\Delta\mathcal{E}_{\rm G}+200\;MeV\;fm^{-3}}$ respectively. The black curve indicates the original EOS. The shaded region shows the acceptance values derived by the GW170817 event~\cite{Abbott-gw170817}.}
    \label{fig:L1L2_GRDF-DD2}
\end{figure*}

\begin{figure*}
    \centering
    \includegraphics[width=\columnwidth]{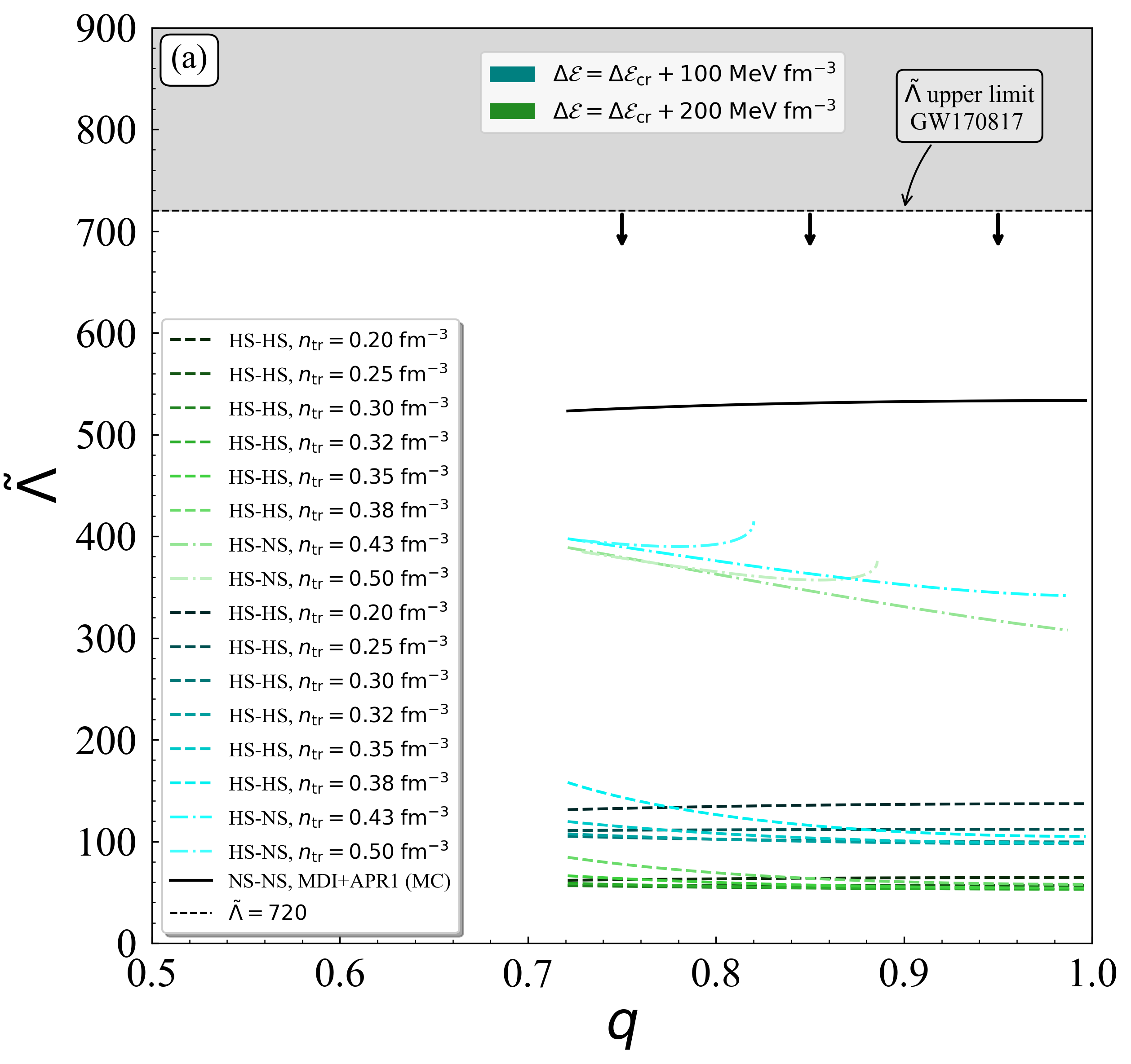}~
    \includegraphics[width=\columnwidth]{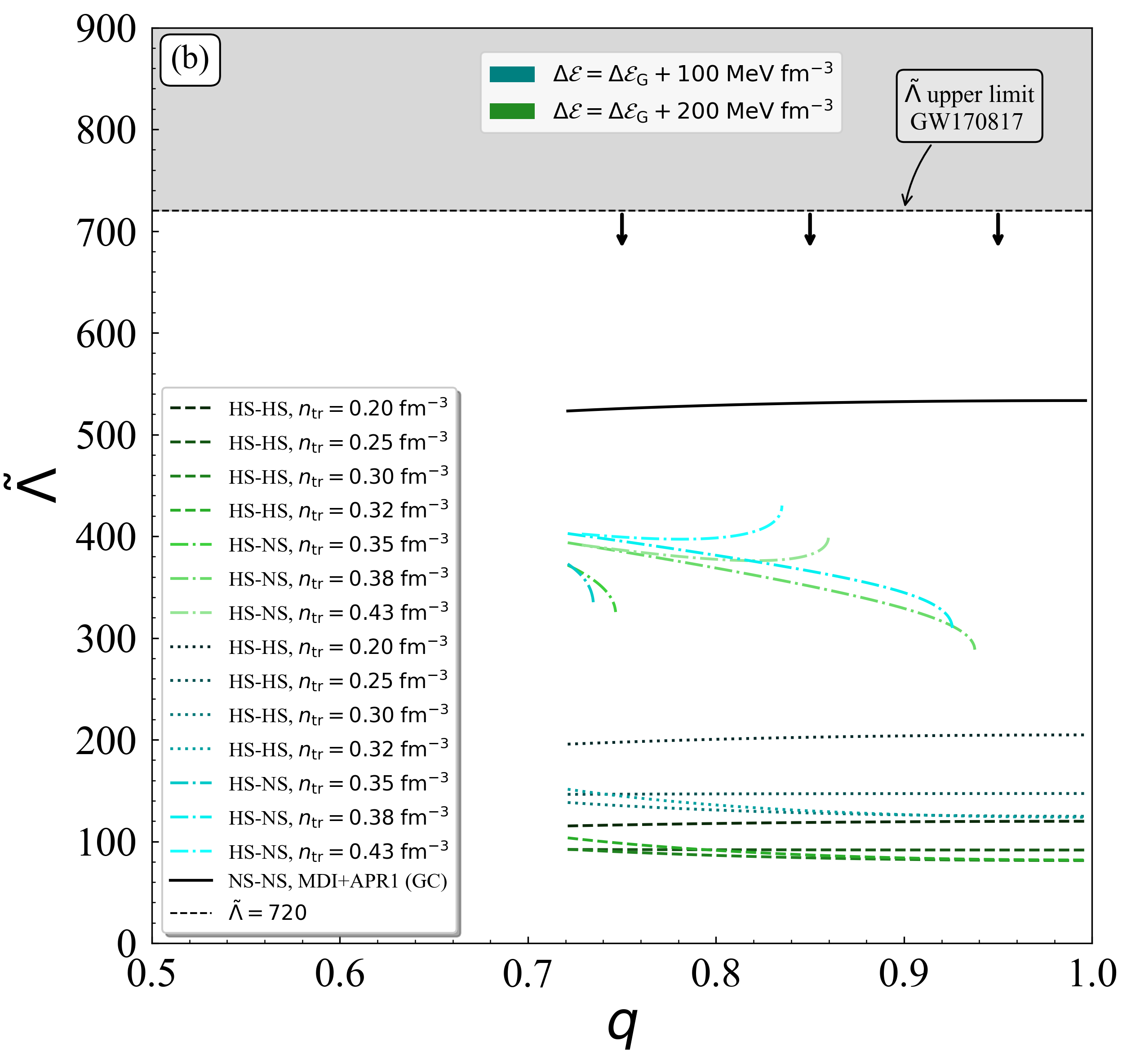}
    \caption{$\tilde{\Lambda}-q$ relation for the MDI+APR1 EOS and a) MC, and b) GC. The blue (green) curves correspond to the $\mathrm{\Delta\mathcal{E}=\Delta \mathcal{E}_{cr}+100\;MeV\;fm^{-3}}$ ($\mathrm{\Delta\mathcal{E}=\Delta \mathcal{E}_{cr}+200\;MeV\;fm^{-3}}$) for the MC, while for the GC correspond to the $\mathrm{\Delta\mathcal{E}=\Delta\mathcal{E}_{\rm G}+100\;MeV\;fm^{-3}}$ and $\mathrm{\Delta\mathcal{E}=\Delta\mathcal{E}_{\rm G}+200\;MeV\;fm^{-3}}$ respectively. The black curve indicates the original EOS. The shaded region shows the acceptance values derived by the GW170817 event~\cite{Abbott-gw170817}.}
    \label{fig:Ltildeq_MDI}
\end{figure*}

\subsection{Tidal Deformability}

Figure~\ref{fig:tidalparams_maxwell_MDI} presents the tidal parameters $k_2$ and $\lambda$ of a single neutron star as a function of its mass, for the MDI+APR1 EOS (MC) and for the cases with $\mathrm{\Delta\mathcal{E}=\Delta \mathcal{E}_{cr}+[100,200]\; MeV\;fm^{-3}}$. The effect of the different $\mathrm{\Delta\mathcal{E}}$ and $\mathrm{n_{tr}}$ can be observed, as they lead to distinct subbranches. In general, the higher value of $\mathrm{\Delta\mathcal{E}}$ lead to softer EOSs across the same bifurcation characterized by the value of transition density $\mathrm{n_{tr}}$.

In Figures~\ref{fig:L1L2_MDI}-\ref{fig:L1L2_GRDF-DD2} we present the $\Lambda_1-\Lambda_2$ space for each EOS and configuration using the observational data of the GW170817 event (orange shaded region)~\cite{Abbott-gw170817}. To be more specific, we considered the three following combinations: Hybrid-Hybrid binary star system (HS-HS), Hybrid-Neutron star system (HS-NS), and finally a Neutron-Neutron one (NS-NS). We notice that we concentrated only in the cases with $\mathrm{\Delta\mathcal{E}=\Delta \mathcal{E}_{cr}+[100,200]\; MeV\;fm^{-3}}$ for MC and $\mathrm{\Delta\mathcal{E}=\Delta \mathcal{E}_{G}+[100,200]\; MeV\;fm^{-3}}$ for GC, because these cases provide more easily a twin star branch on the EOS. The blue curves correspond to $\mathrm{\Delta\mathcal{E}=\Delta \mathcal{E}_{cr}+100\; MeV\;fm^{-3}}$ ($\mathrm{\Delta\mathcal{E}=\Delta \mathcal{E}_{G}+100\; MeV\;fm^{-3}}$) for MC (GC), while the green ones to $\mathrm{\Delta\mathcal{E}=\Delta \mathcal{E}_{cr}+200\; MeV\;fm^{-3}}$ ($\mathrm{\Delta\mathcal{E}=\Delta \mathcal{E}_{G}+200\; MeV\;fm^{-3}}$) for MC (GC), respectively. In all diagrams, the dashed curves correspond to the HS-HS case, the dash-dotted curves to the HS-NS case, and the solid curves to the NS-NS case.

In more detail, for the MDI+APR1 EOS and MC we used the values a) of $\mathrm{n_{tr}}=[0.20,0.25,0.30,0.32,0.35,0.38] \; \mathrm{fm^{-3}}$ for the HS-HS case, and b) of $\mathrm{n_{tr}}=[0.43,0.50] \; \mathrm{fm^{-3}}$ for the HS-NS, while for the MDI+APR1 EOS and GC we used a) $\mathrm{n_{tr}}=[0.20,0.25,0.30,0.32] \; \mathrm{fm^{-3}}$ for the HS-HS case, and b) $\mathrm{n_{tr}}=[0.35,0.38,0.43] \; \mathrm{fm^{-3}}$ for the HS-NS case. Moving on to the GRDF-DD2 EOS, we notice that for the MC we used a) $\mathrm{n_{tr}}=[0.20,0.25,0.30,0.32] \; \mathrm{fm^{-3}}$ for the HS-HS case, and b) $\mathrm{n_{tr}}=[0.35,0.38] \; \mathrm{fm^{-3}}$ for the HS-NS case, while for the GC we used a) $\mathrm{n_{tr}}=[0.20,0.25] \; \mathrm{fm^{-3}}$ for the HS-HS case, and b) $\mathrm{n_{tr}}=[0.30,0.32] \; \mathrm{fm^{-3}}$ for the HS-NS case. We remark that even that we kept the region of the component masses identical to the GW170817 observation for the majority of the cases of those hypothetical binary star systems, in some cases we modified and restricted the mass range for computational reasons.

In all cases, the HS-HS lead to smaller values of $\Lambda$, therefore to softer EOS, in accordance to the observational data of GW170817. The HS-NS case of a binary star system lies also inside the shaded region provided from LIGO, but in intermediate values of $\Lambda$. The NS-NS is the case where the two component stars correspond to the original EOS that we used in each case (the solid black line in M-R diagrams). As one can observe, the MC provides more cases for the HS-HS scenario compared to the GC, for both EOSs that we used in our study. In addition, the MDI+APR1 EOS for the NS-NS case lies inside the estimation of GW170817 nevertheless the combination that we considered, on the contrary of the GRDF-DD2 EOS, in which the  NS-NS cases lie outside the estimated region.

\begin{figure*}
    \centering
    \includegraphics[width=\columnwidth]{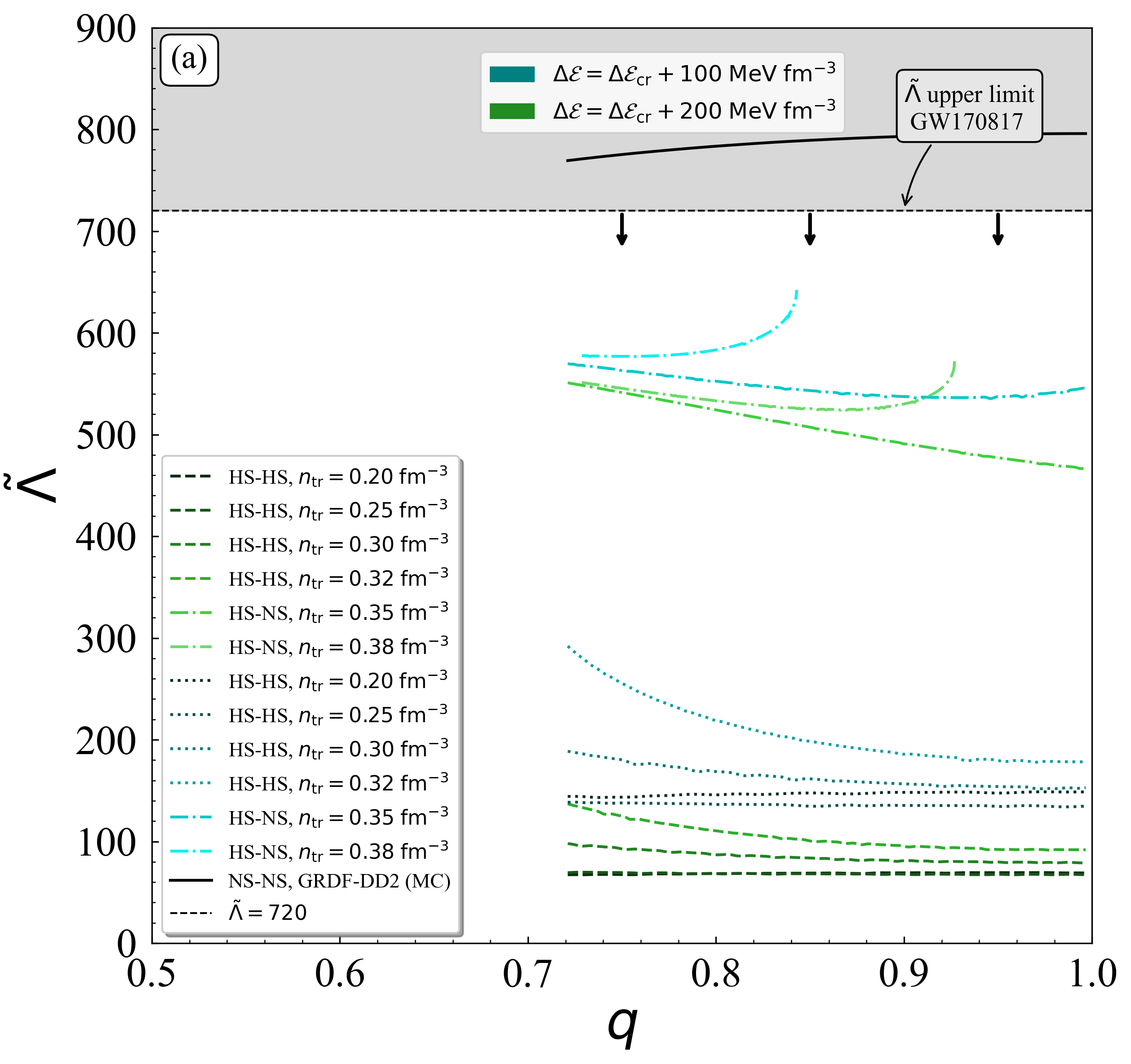}~
    \includegraphics[width=\columnwidth]{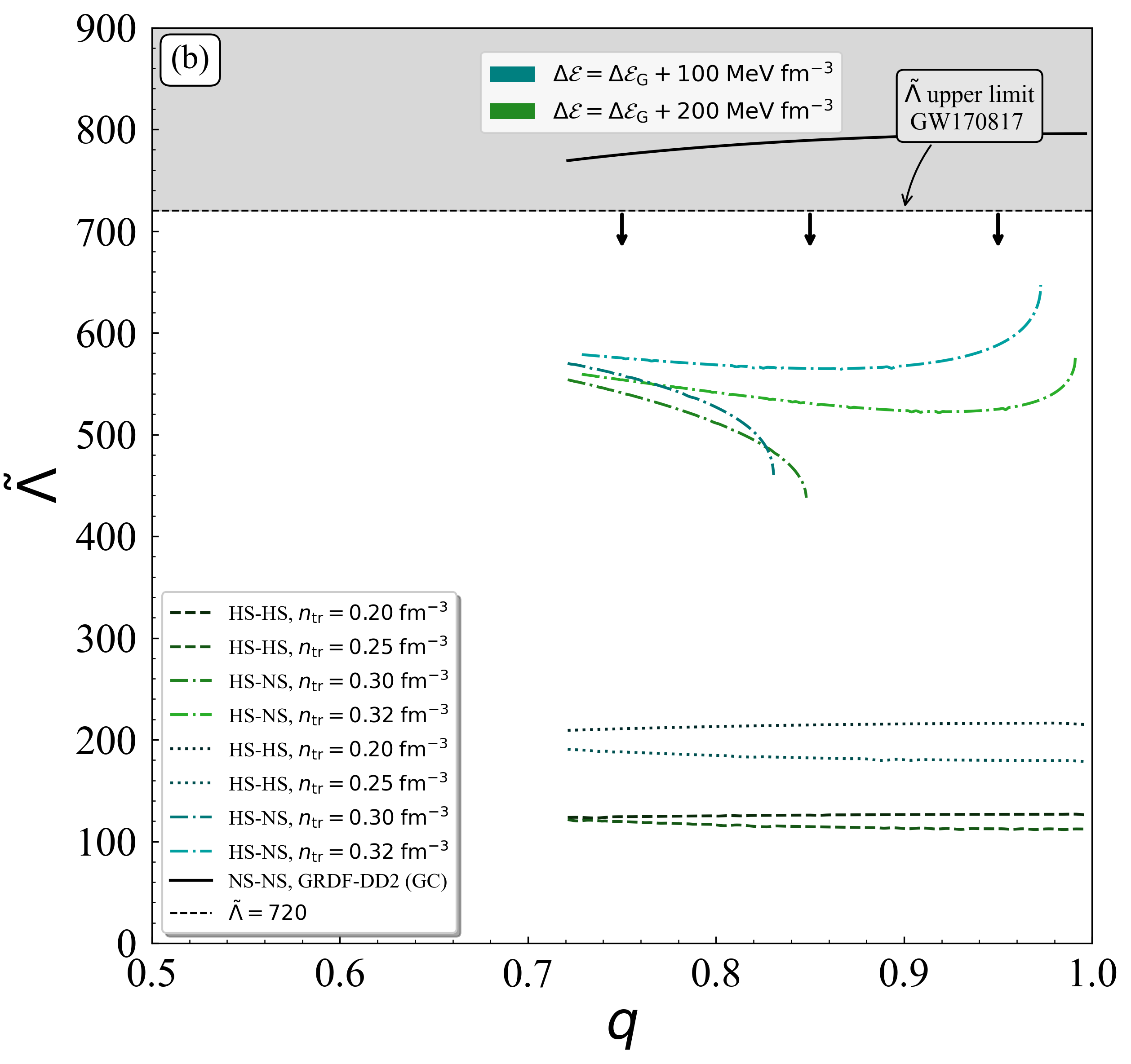}
    \caption{$\tilde{\Lambda}-q$ relation for the GRDF-DD2 EOS and a) MC, and b) GC. The blue (green) curves correspond to the $\mathrm{\Delta\mathcal{E}=\Delta \mathcal{E}_{cr}+100\;MeV\;fm^{-3}}$ ($\mathrm{\Delta\mathcal{E}=\Delta \mathcal{E}_{cr}+200\;MeV\;fm^{-3}}$) for the MC, while for the GC correspond to the $\mathrm{\Delta\mathcal{E}=\Delta\mathcal{E}_{\rm G}+100\;MeV\;fm^{-3}}$ and $\mathrm{\Delta\mathcal{E}=\Delta\mathcal{E}_{\rm G}+200\;MeV\;fm^{-3}}$ respectively. The black curve indicates the original EOS. The shaded region shows the acceptance values derived by the GW170817 event~\cite{Abbott-gw170817}.}
    \label{fig:Ltildeq_GRDF-DD2}
\end{figure*}

\begin{figure*}
    \centering
    \includegraphics[width=\columnwidth]{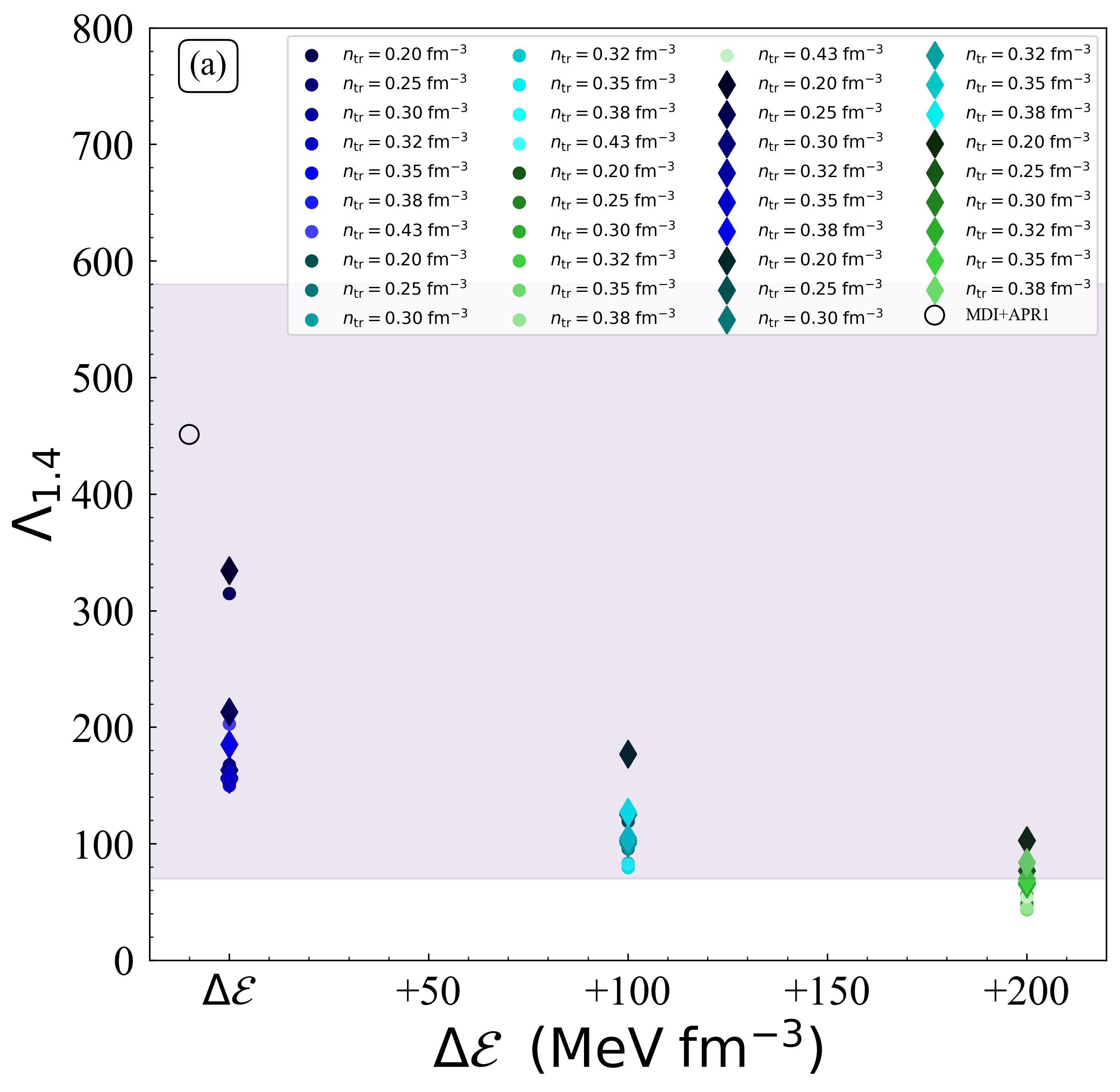}~
    \includegraphics[width=\columnwidth]{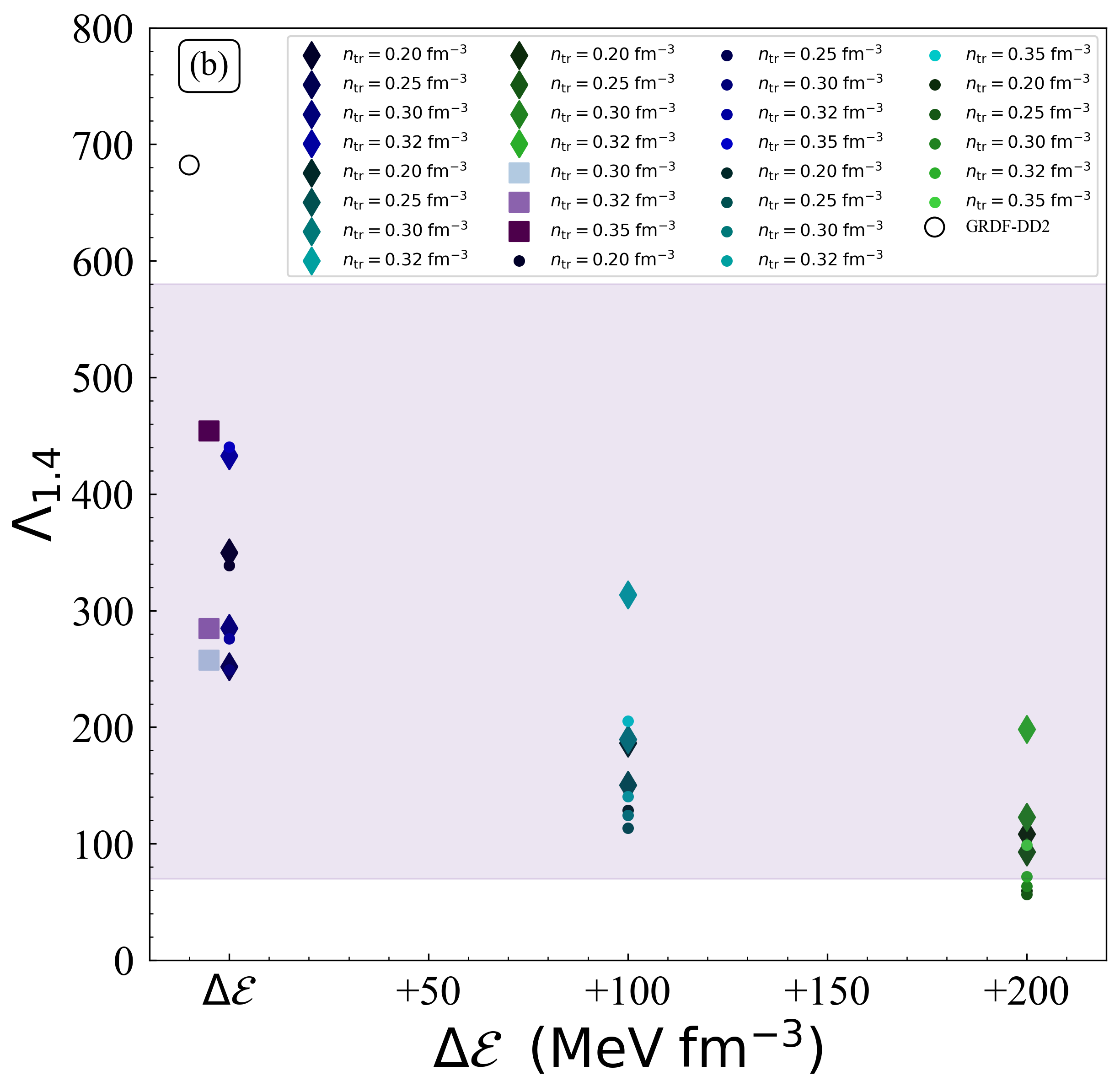}
      \caption{$\Lambda_{1.4}-\Delta\mathcal{E}$ diagram for a single $1.4M_\odot$ neutron star for a) the MDI+APR1 (left panel), and b) GRDF-DD2 (right panel) EOS for both MC (circles) and GC (diamonds). The square marks correspond to the $\mathrm{\Delta\mathcal{E}=\Delta\mathcal{E}_{\rm cr}-5\;MeV\;fm^{-3}}$ case under MC and only for the GRDF-DD2 EOS, the dark blue marks correspond to the $\Delta \mathcal{E}_{\rm cr}$ ($\Delta \mathcal{E}_{\rm G}$) case, the light blue marks correspond to $\mathrm{\Delta\mathcal{E}=\Delta\mathcal{E}_{\rm cr}+100\;MeV\;fm^{-3}}$ ($\mathrm{\Delta\mathcal{E}=\Delta\mathcal{E}_{\rm G}+100\;MeV\;fm^{-3}}$), and the green ones correspond to $\mathrm{\Delta\mathcal{E}=\Delta\mathcal{E}_{\rm cr}+200\;MeV\;fm^{-3}}$ ($\mathrm{\Delta\mathcal{E}=\Delta\mathcal{E}_{\rm G}+200\;MeV\;fm^{-3}}$) for the MC (GC) respectively. The empty colored circle in each panel indicates the original EOS. The light purple shaded region corresponds to the GW170817 event~\cite{Abbott-2}.}
    \label{fig:L1_4_DE}
\end{figure*}

In Figures~\ref{fig:Ltildeq_MDI}-\ref{fig:Ltildeq_GRDF-DD2} we show the $\tilde{\Lambda}-q$ relation, by using the upper limit on $\tilde{\Lambda}$, provided by the GW170817 event~\cite{Abbott-gw170817}. The curves and colors are the similar to those of $\Lambda_1-\Lambda_2$ diagrams. As one can observe, the HS-HS case for both EOSs and constructions lead to much lower values of $\tilde{\Lambda}$. Hence, a possible lower limit on $\tilde{\Lambda}$ would be very useful to restrict the lower values of $\tilde{\Lambda}$, leading to constraints at least on the HS-HS case. Moreover, as we move to higher values of $\mathrm{\Delta\mathcal{E}}$, all the curves for both HS-HS and HS-NS scenarios are shifted to lower values of $\tilde{\Lambda}$, meaning that the increment of $\mathrm{\Delta\mathcal{E}}$ has as a result softer EOSs. This behavior was expected if we recall the effect of higher values of $\mathrm{\Delta\mathcal{E}}$ on the EOSs (see M-R diagrams). Between the two constructions, and for both EOSs, the MC provides higher number of HS-HS cases compared to the GC. Also, across the same EOS and $\mathrm{n_{tr}}$, the MC leads to a bit of lower values of $\tilde{\Lambda}$ compared to the GC. Therefore, for low mass events, such as the GW170817, MC is more suitable.

The need for a lower limit on $\tilde{\Lambda}$ that we described before, led us to the exploit of the constrained value of the dimensionless tidal deformability for a single $1.4M_\odot$ neutron star, derived by the study of the GW170817 event. In Figure~\ref{fig:L1_4_DE} we show the relation between $\Lambda_{1.4}$ and $\mathrm{\Delta\mathcal{E}}$. We notice that we

\begin{figure}[H]
    \centering
    \includegraphics[width=\columnwidth]{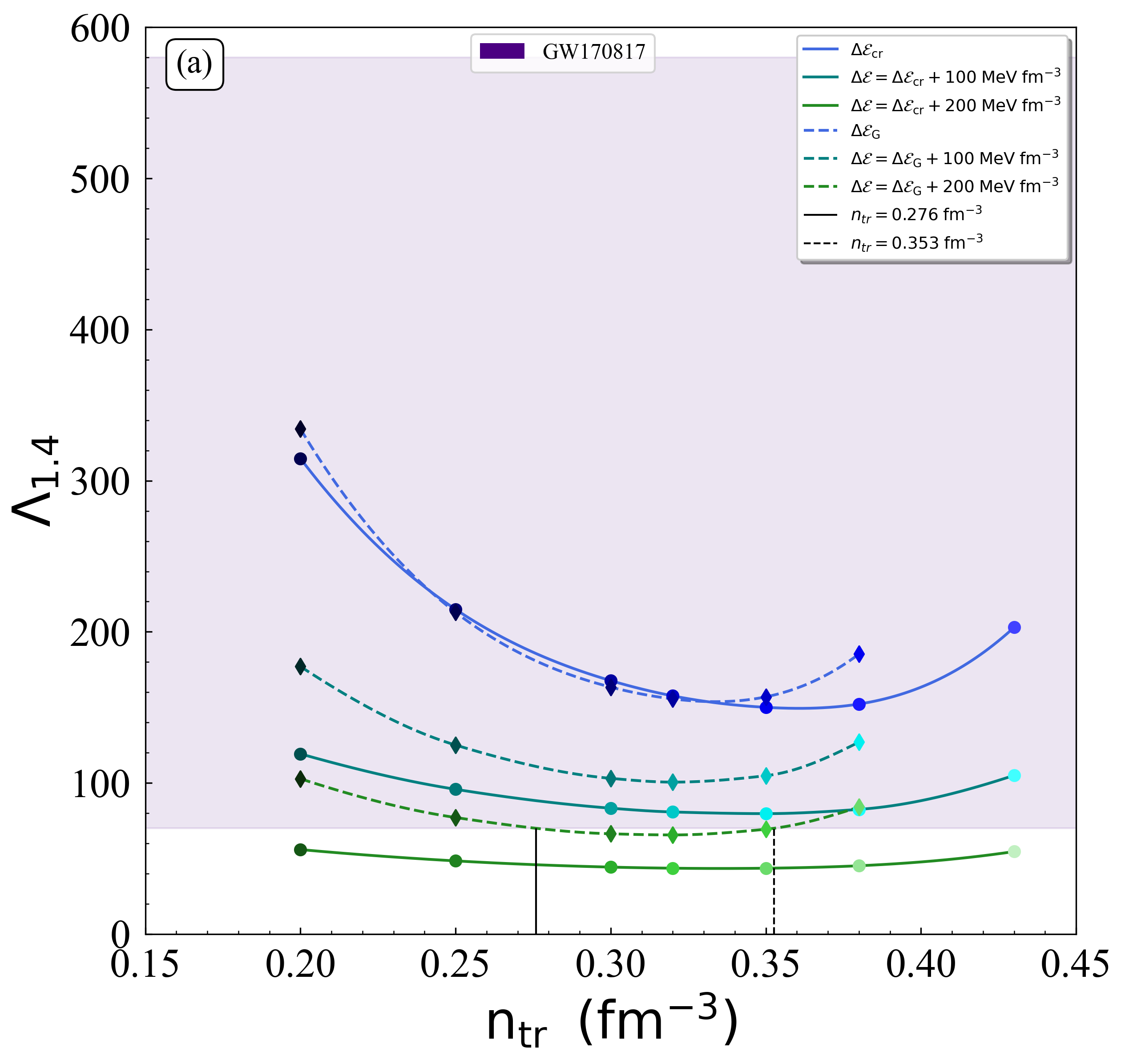}
    
    \includegraphics[width=\columnwidth]{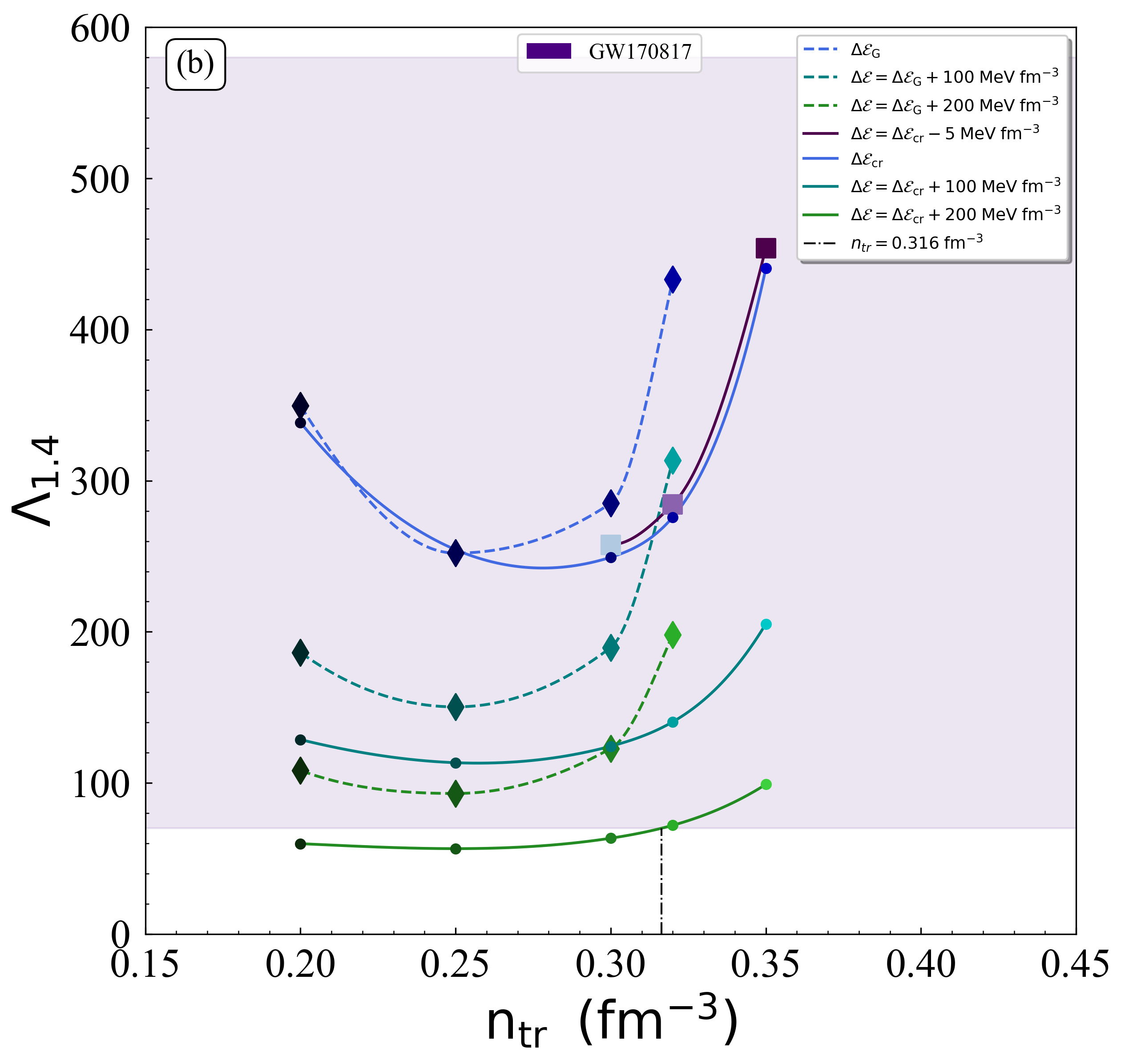}
      \caption{$\Lambda_{1.4}-n_{tr}$ diagram for a single $1.4M_\odot$ neutron star for a) the MDI+APR1 (left panel), and b) GRDF-DD2 (right panel) EOS for both MC (solid lines) and GC (dashed lines). The dark purple line corresponds to the $\mathrm{\Delta\mathcal{E}=\Delta\mathcal{E}_{\rm cr}-5\;MeV\;fm^{-3}}$ case under MC and only for the GRDF-DD2 EOS, the dark blue lines correspond to the $\Delta \mathcal{E}_{\rm cr}$ ($\Delta \mathcal{E}_{\rm G}$) case, the light blue lines correspond to $\mathrm{\Delta\mathcal{E}=\Delta\mathcal{E}_{\rm cr}+100\;MeV\;fm^{-3}}$ ($\mathrm{\Delta\mathcal{E}=\Delta\mathcal{E}_{\rm G}+100\;MeV\;fm^{-3}}$), and the green ones correspond to $\mathrm{\Delta\mathcal{E}=\Delta\mathcal{E}_{\rm cr}+200\;MeV\;fm^{-3}}$ ($\mathrm{\Delta\mathcal{E}=\Delta\mathcal{E}_{\rm G}+200\;MeV\;fm^{-3}}$) for the MC (GC) respectively. The light purple shaded region corresponds to the GW170817 event~\cite{Abbott-2}.}
    \label{fig:L1_4_ntr}
\end{figure}

\noindent used only those values of $\mathrm{n_{tr}}$ that provide a separate branch from the original EOS. As one can observe, as we move from the $\Delta \mathcal{E}_{\rm cr}$ (and $\Delta \mathcal{E}_{\rm G}$ for the GC) to higher values of $\mathrm{\Delta\mathcal{E}}$ the variation between the marks decreases. In addition, all the marks that correspond to the GC (diamonds) predict higher values of $\Lambda_{1.4}$ compared to the MC, for both EOSs. We notice also that only for the $\mathrm{\Delta\mathcal{E}=\Delta \mathcal{E}_{cr}+200\;MeV\;fm^{-3}}$ and $\mathrm{\Delta\mathcal{E}=\Delta \mathcal{E}_{G}+200\;MeV\;fm^{-3}}$ there is a violation of the accepted region. Moreover, in Figure~\ref{fig:L1_4_DE}(b), the square points indicate the $\mathrm{\Delta\mathcal{E}=\Delta \mathcal{E}_{cr}-5\;MeV\;fm^{-3}}$ case under MC and for the GRDF-DD2 EOS. As one can observe, as we move to lower values compared to $\Delta \mathcal{E}_{cr}$, the variation increases, which is in accordance with the behavior that we described before.

Furthermore, in order to shed more light on which specific cases should be excluded, we studied the relation between $\Lambda_{1.4}$ and $\mathrm{n_{tr}}$. In Figure~\ref{fig:L1_4_ntr} we present the aforementioned relation. As a first remark, the MC provides in all cases one more value of $\mathrm{n_{tr}}$ compared to the GC. The GC shifts the curves (dashed) to higher values compared to the relevant curves of MC (solid). In addition, as we move to higher values of $\mathrm{\Delta\mathcal{E}}$ the curves are shifted to lower values of $\Lambda_{1.4}$.

The left panel of the Figure~\ref{fig:L1_4_ntr} corresponds to the MDI+APR1 EOS. The curves that correspond to the $\Delta \mathcal{E}_{\rm cr}$ and $\mathrm{\Delta\mathcal{E}=\Delta \mathcal{E}_{cr}+100\;MeV\;fm^{-3}}$ lie inside the estimated region. The same holds for $\Delta\mathcal{E_{\rm G}}$, and $\mathrm{\Delta\mathcal{E}=\Delta \mathcal{E}_{G}+100\;MeV\;fm^{-3}}$. On the other hand, the solid green curve, which corresponds to the MC with $\mathrm{\Delta\mathcal{E}=\Delta \mathcal{E}_{cr}+200\;MeV\;fm^{-3}}$ is excluded. But if we apply the GC, the curve is shifted upwards (dashed green curve), with only a part being outside of the estimated region. Especially, this part is between $\mathrm{n_{tr}=0.276\; fm^{-3}}$ and $\mathrm{n_{tr}=0.353\; fm^{-3}}$. Hence, not only the kind of construction we choose has a significant role, but the exact value of the transition density affects the final output. Therefore, a further understanding and possible constraints on the transition density $\mathrm{n_{tr}}$ are necessary to shed more light on the twin star hypothesis.

The right panel of the Figure~\ref{fig:L1_4_ntr} corresponds to the GRDF-DD2 EOS. The curves that correspond to the $\Delta \mathcal{E}_{\rm cr}$ and $\mathrm{\Delta\mathcal{E}=\Delta \mathcal{E}_{cr}+100\;MeV\;fm^{-3}}$ lie inside the estimated region. Only the green solid curve which corresponds to the MC with $\mathrm{\Delta\mathcal{E}=\Delta \mathcal{E}_{cr}+200\;MeV\;fm^{-3}}$ lies outside up to $\mathrm{n_{tr}=0.316\; fm^{-3}}$, meaning that above this value even this EOS could be acceptable. All the curves that correspond to the GC lie inside the estimated region. As we mentioned above, the construction and the transition density affect importantly the behavior of the curves. We notice that the purple line indicates the $\mathrm{\Delta\mathcal{E}=\Delta \mathcal{E}_{cr}-5\;MeV\;fm^{-3}}$ case under MC. As we have already noticed, as we move to lower values of $\Delta\mathcal{E}$ compared to $\Delta\mathcal{E}_{\rm cr}$, the EOS becomes stiffer, therefore in this diagram the curve is shifted sligtly to higher values of $\Lambda_{1.4}$.

\begin{figure}
    \centering
    \includegraphics[width=\columnwidth]{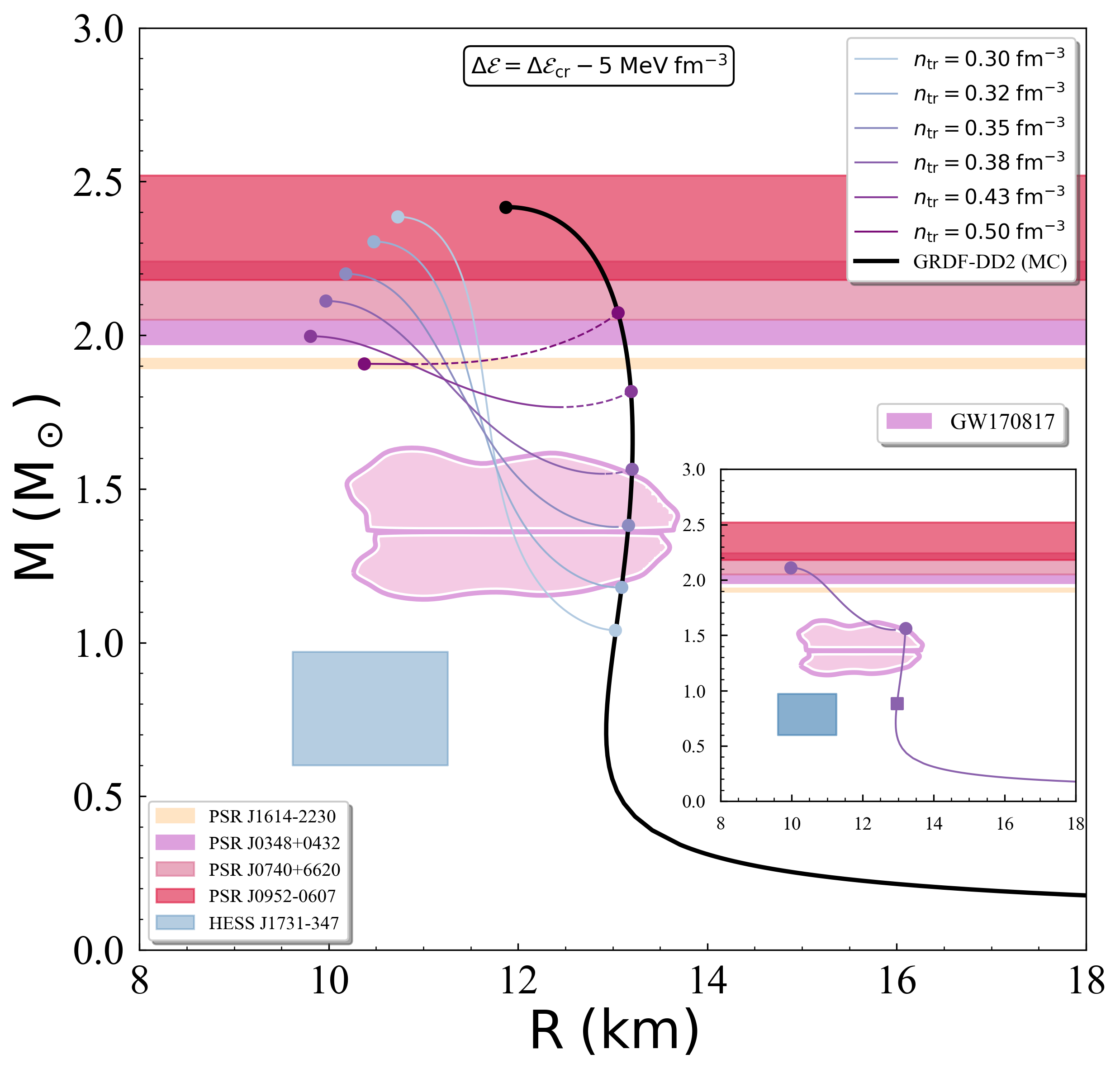}
    \caption{Mass vs Radius for the GRDF-DD2 EOS under MC and for $\Delta \mathcal{E} = \Delta \mathcal{E}_{\rm cr} - 5~{\rm MeV~fm^{-3}}$. The dashed part of curves indicates their unstable region. The black curve indicates the original EOS. The shaded regions from bottom to top represent the HESS J1731-347 remnant~\cite{Doroshenko-2022}, GW170817 event~\cite{Abbott-gw170817}, PSR J1614–2230~\cite{Arzoumanian-2018}, PSR J0348+0432~\cite{Antoniadis-2013}, PSR J0740+6620~\cite{Cromartie-2020}, and PSR J0952-0607~\cite{Romani-2022} pulsar observations for possible maximum mass. The inset plot indicates the case with $\mathrm{n_{tr}=0.38\;fm^{-3}}$, in which the square point represents the phase transition.}
    \label{fig:MR_nsl_5_GRDF-DD2_MC}
\end{figure}

In   Figure~\ref{fig:MR_nsl_5_GRDF-DD2_MC}  we show the M-R diagram  for the GRDF-DD2 EOS under MC and for $\Delta \mathcal{E} = \Delta \mathcal{E}_{\rm cr} - 5~{\rm MeV~fm^{-3}}$, which corresponds to a prediction under the Seidov limit. It is interesting that  this case predicts the existence of a second stable branch and consequently, the existence of twin star (and indeed within the predictions of GW170817 event~\cite{Abbott-gw170817}). However, this case cannot predict the existence of the compact object  of the remnant HESS J1731-347~\cite{Doroshenko-2022}. These results confirm those already presented  in Figures~\ref{fig:MR_maxwell_MDI}-\ref{fig:MR_gibbs_GRDF-DD2}, that is, hybrid EOSs with a large energy density  gap are preferable for explaining the existence of the above compact object.

\subsection{Rotational frequency at 709 Hz and the PSR J0952-0607 pulsar}
\begin{figure*}
    \centering
    \includegraphics[width=\textwidth]{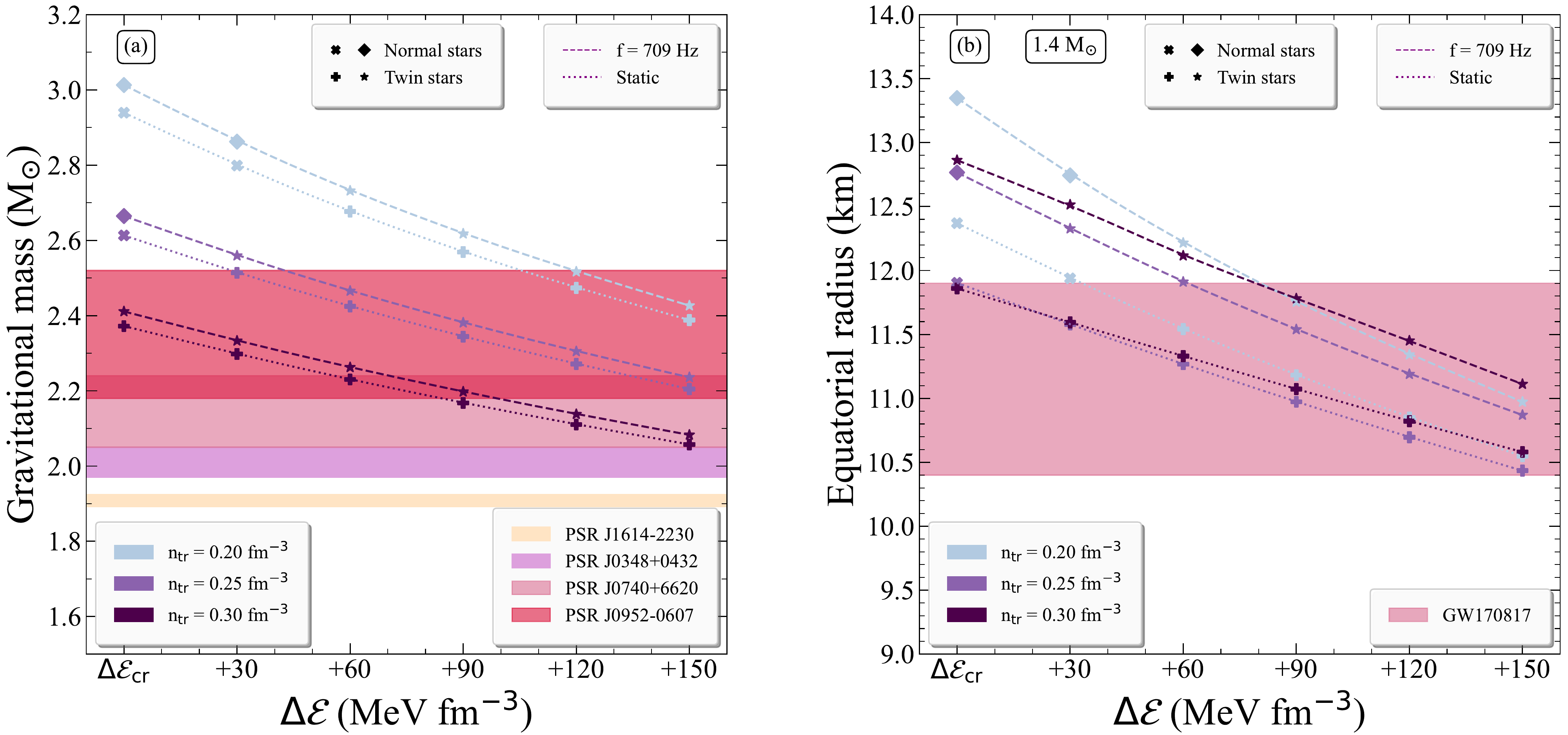}
    \caption{(a) Gravitational mass as a function of the energy jump for transition densities in the range $[0.2,0.3]~{\rm fm^{-3}}$ at the maximum mass configuration. The shaded regions from bottom to top represent the PSR J1614–2230~\cite{Arzoumanian-2018}, PSR J0348+0432~\cite{Antoniadis-2013}, PSR J0740+6620~\cite{Cromartie-2020}, and PSR J0952-0607~\cite{Romani-2022} pulsar observations for possible maximum mass. (b) Equatorial radius as a function of the energy jump for transition densities in the range $[0.2,0.3]~{\rm fm^{-3}}$ at the $\rm 1.4~M_{\odot}$ configuration. The shaded region represents the constraints extracted through GW170817 event~\cite{Capano-2020}. Normal neutron stars are presented with the diamonds and crosses, corresponding to the rotating at 709 Hz configuration and the non-rotating one, respectively, while twin stars, use the stars and plus signs. Both figures correspond to the Maxwell construction method.}
    \label{fig:mass_energy_rotation_maxwell}
\end{figure*}

\begin{figure*}
    \centering
    \includegraphics[width=\textwidth]{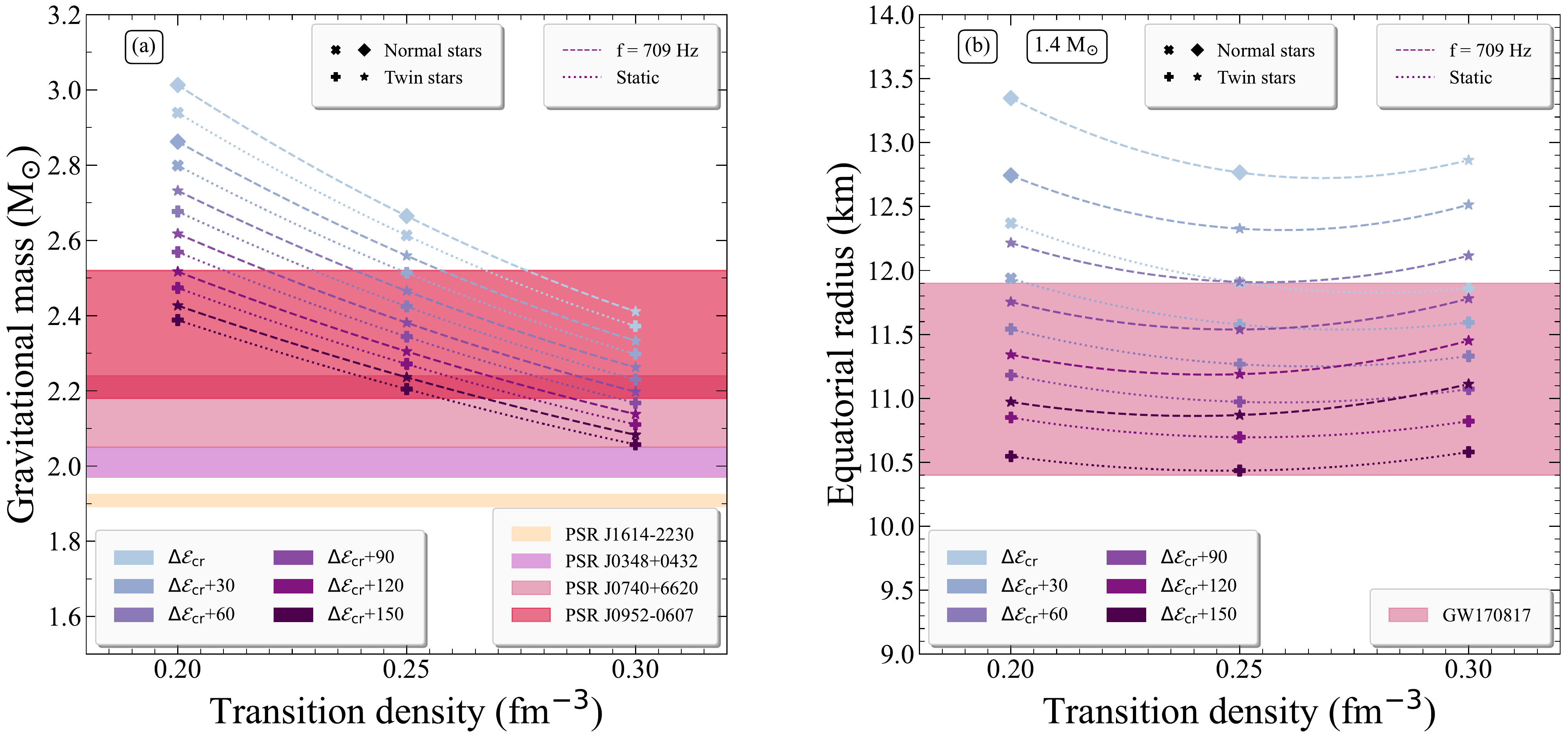}
    \caption{(a) Gravitational mass as a function of the transition density for energy jumps in the range $[\Delta \mathcal{E}_{\rm cr}, \Delta \mathcal{E}_{\rm cr}+150]~{\rm MeV~fm^{-3}}$ at the maximum mass configuration. The shaded regions from bottom to top represent the PSR J1614–2230~\cite{Arzoumanian-2018}, PSR J0348+0432~\cite{Antoniadis-2013}, PSR J0740+6620~\cite{Cromartie-2020}, and PSR J0952-0607~\cite{Romani-2022} pulsar observations for possible maximum mass. (b) Equatorial radius as a function of the transition density for energy jumps in the range $[\Delta \mathcal{E}_{\rm cr}, \Delta \mathcal{E}_{\rm cr}+150]~{\rm MeV~fm^{-3}}$ at the $\rm 1.4~M_{\odot}$ configuration. The shaded region represents the constraints extracted through GW170817 event~\cite{Capano-2020}. Normal neutron stars are presented with the diamonds and crosses, corresponding to the rotating at 709 Hz configuration and the non-rotating one, respectively, while twin stars, use the stars and plus signs. Both figures correspond to the Maxwell construction method.}
    \label{fig:mass_density_rotation_maxwell}
\end{figure*}

\begin{figure*}
    \centering
    \includegraphics[width=\textwidth]{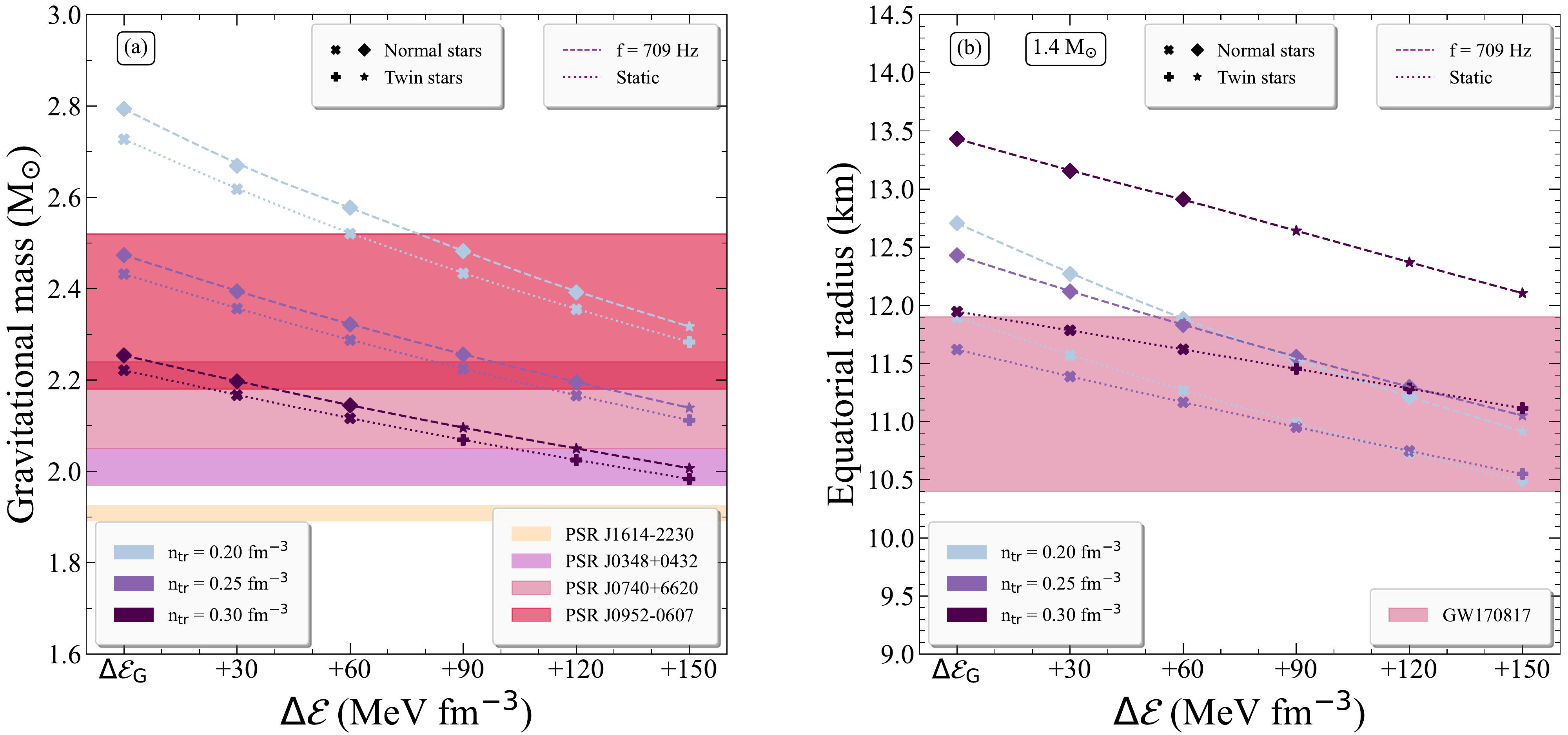}
    \caption{(a) Gravitational mass as a function of the energy increase for transition densities in the range $[0.2,0.3]~{\rm fm^{-3}}$ at the maximum mass configuration. The shaded regions from bottom to top represent the PSR J1614–2230~\cite{Arzoumanian-2018}, PSR J0348+0432~\cite{Antoniadis-2013}, PSR J0740+6620~\cite{Cromartie-2020}, and PSR J0952-0607~\cite{Romani-2022} pulsar observations for possible maximum mass. (b) Equatorial radius as a function of the energy increase for transition densities in the range $[0.2,0.3]~{\rm fm^{-3}}$ at the $\rm 1.4~M_{\odot}$ configuration. The shaded region represents the constraints extracted through GW170817 event~\cite{Capano-2020}. Normal neutron stars are presented with the diamonds and crosses, corresponding to the rotating at 709 Hz configuration and the non-rotating one, respectively, while twin stars, use the stars and plus signs. Both figures correspond to the Gibbs construction method.}
    \label{fig:mass_energy_rotation_gibbs}
\end{figure*}

\begin{figure*}
    \centering
    \includegraphics[width=\textwidth]{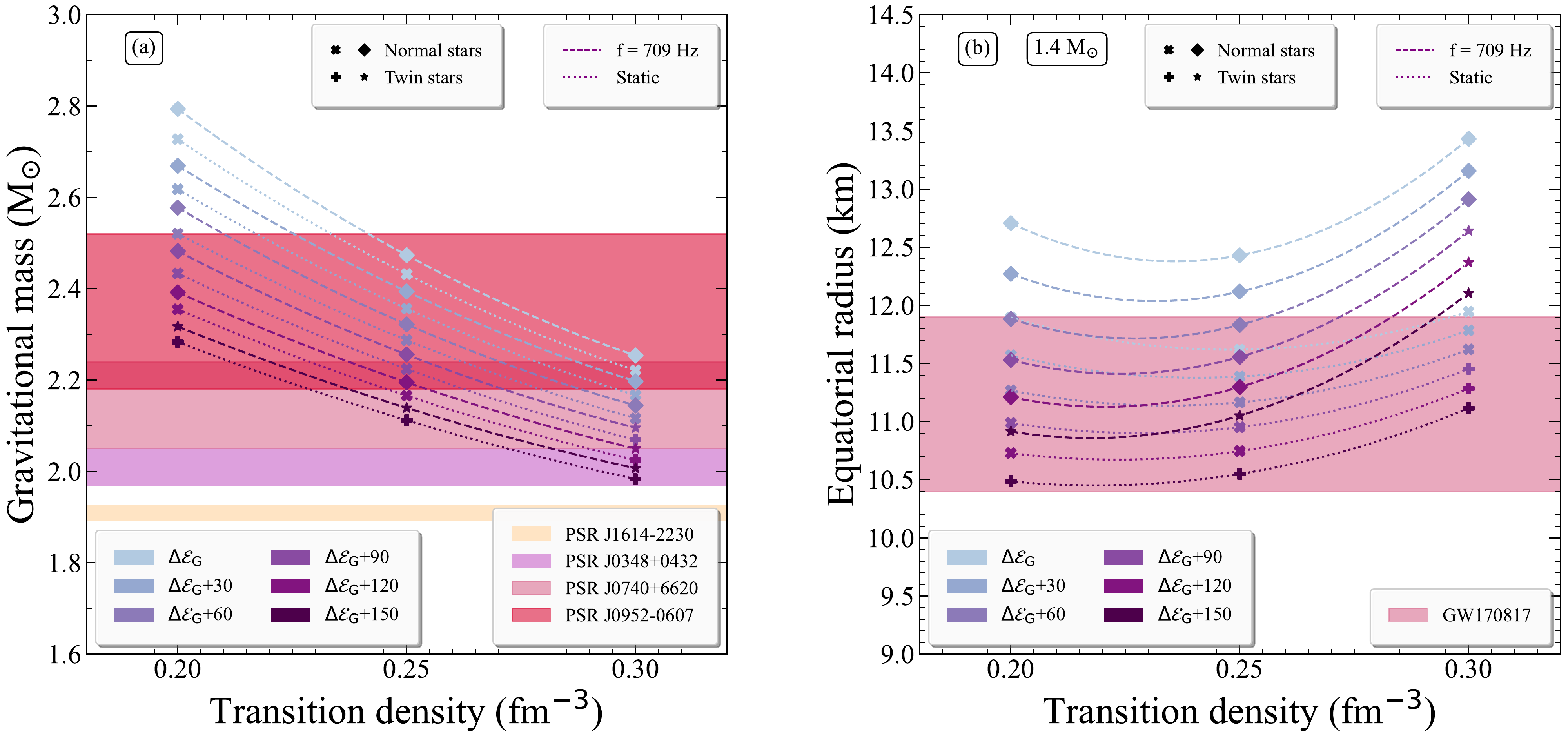}
    \caption{(a) Gravitational mass as a function of the transition density for energy increases in the range $[\Delta \mathcal{E}_{\rm G}, \Delta \mathcal{E}_{\rm G}+150]~{\rm MeV~fm^{-3}}$ at the maximum mass configuration. The shaded regions from bottom to top represent the PSR J1614–2230~\cite{Arzoumanian-2018}, PSR J0348+0432~\cite{Antoniadis-2013}, PSR J0740+6620~\cite{Cromartie-2020}, and PSR J0952-0607~\cite{Romani-2022} pulsar observations for possible maximum mass. (b) Equatorial radius as a function of the transition density for energy increases in the range $[\Delta \mathcal{E}_{\rm G}, \Delta \mathcal{E}_{\rm G}+150]~{\rm MeV~fm^{-3}}$ at the $\rm 1.4~M_{\odot}$ configuration. The shaded region represents the constraints extracted through GW170817 event~\cite{Capano-2020}. Normal neutron stars are presented with the diamonds and crosses, corresponding to the rotating at 709 Hz configuration and the non-rotating one, respectively, while twin stars, use the stars and plus signs. Both figures correspond to the Gibbs construction method.}
    \label{fig:mass_density_rotation_gibbs}
\end{figure*}

\begin{figure}
    \centering
    \includegraphics[width=\columnwidth]{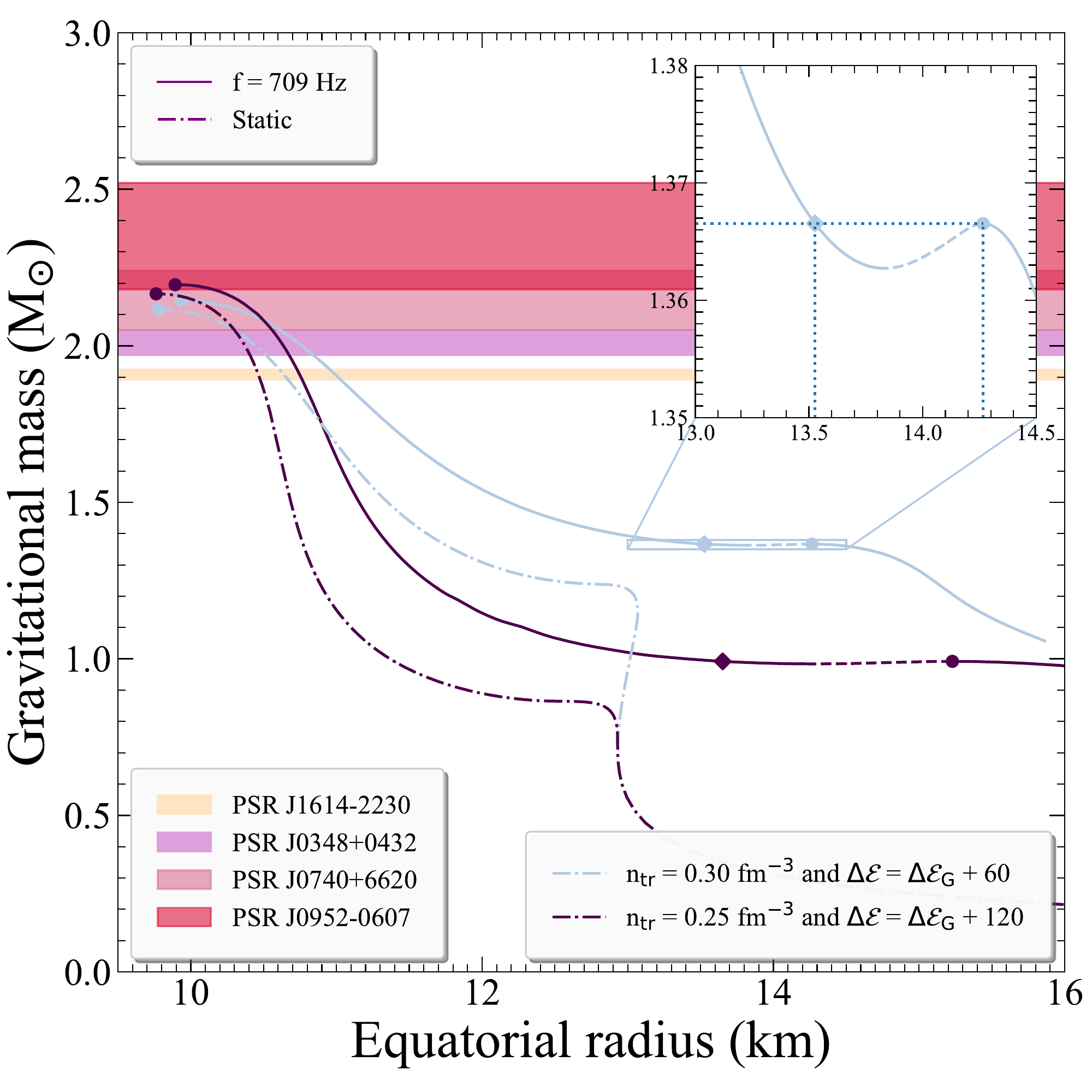}
    \caption{Gravitational mass as a function of the equatorial radius for two representative cases of the GC: (a) $n_{\rm tr} = 0.25~{\rm fm^{-3}}$ and $\Delta \mathcal{E} = \Delta \mathcal{E}_{\rm G} + 120~{\rm MeV~fm^{-3}}$ and (b) $n_{\rm tr} = 0.3~{\rm fm^{-3}}$ and $\Delta \mathcal{E} = \Delta \mathcal{E}_{\rm G} + 60~{\rm MeV~fm^{-3}}$ . The circles represent the maximum mass configuration, while the diamonds correspond to the twin stars, assuming mass equal to the first maximum mass configuration. The inset figure indicates with accuracy the twin star branch for the case $n_{\rm tr} = 0.3~{\rm fm^{-3}}$ and $\Delta \mathcal{E} = \Delta \mathcal{E}_{\rm G} + 60~{\rm MeV~fm^{-3}}$. The dash-dotted lines represent the non-rotating configuration, while the solid lines correspond to the 709 Hz configuration. To guide the eye, the unstable region is marked with the dashed lines. The shaded regions from bottom to top represent the PSR J1614–2230~\cite{Arzoumanian-2018}, PSR J0348+0432~\cite{Antoniadis-2013}, PSR J0740+6620~\cite{Cromartie-2020}, and PSR J0952-0607~\cite{Romani-2022} pulsar observations for possible maximum mass.}
    \label{fig:mr_gibbs_rot}
\end{figure}

The simultaneous observation of mass and frequency of the PSR J0952-0607 pulsar~\cite{Romani-2022} opens a new window in studying the rotating configuration for neutron stars and the possible existence of rotating twin stars. In Figures~\ref{fig:mass_energy_rotation_maxwell} and~\ref{fig:mass_energy_rotation_gibbs} we present three cases of transition density, $n_{\rm tr} = [0.2, 0.25, 0.3]~{\rm fm^{-3}}$, for both the non-rotating and rotating at 709 Hz configuration, with MC and GC respectively. Furthermore, we extended the study in a region of the energy jump, starting from the $\Delta \mathcal{E}_{\rm cr}$ for MC and $\Delta \mathcal{E}_{\rm G}$ for GC and reaching values up to +150 ${\rm MeV~fm^{-3}}$. In particular, while in Figure~\ref{fig:mass_energy_rotation_maxwell}(a) there are numerous cases that exist either in the mass limit of the rotating neutron star or in the non-rotating one, there are only three cases where both configurations meet the mass limits simultaneously~\cite{Arzoumanian-2018,Antoniadis-2013,Cromartie-2020,Romani-2022}. Specifically, (a) the $n_{\rm tr} = 0.25~{\rm fm^{-3}}$ and $\Delta \mathcal{E} =\Delta \mathcal{E}_{\rm cr} + 150~{\rm MeV~fm^{-3}}$, (b) the $n_{\rm tr} = 0.3~{\rm fm^{-3}}$ and $\Delta \mathcal{E} = \Delta \mathcal{E}_{\rm cr} + 60~{\rm MeV~fm^{-3}}$, and (c) the $n_{\rm tr} = 0.3~{\rm fm^{-3}}$ and $\Delta \mathcal{E} = \Delta \mathcal{E}_{\rm cr} + 90~{\rm MeV~fm^{-3}}$. In addition, the mentioned cases also fulfil the non-rotating radius limit~\cite{Capano-2020}, as shown in Figure~\ref{fig:mass_energy_rotation_maxwell}(b), extracted from the GW170817 event~\cite{Capano-2020} for the $\rm 1.4~M_{\odot}$ configuration. In the latter, we also present, for completeness and comparison, the corresponding radius at 709 Hz. However, in Figure~\ref{fig:mass_energy_rotation_maxwell}(a), there are numerous cases that can describe solely the 709 Hz configuration. It has to be noted that the mentioned cases are eligible for the description of the PSR J0952-0607~\cite{Romani-2022} pulsar, since we do not know the corresponding non-rotating limit for the mass. On the contrary, in Figure~\ref{fig:mass_energy_rotation_gibbs}(a) there are only a few cases that represent twin stars, which nearly all of them (except one) meet the current limits of the gravitational mass only at the non-rotating configuration. In addition, their radius is close to the limits, or even fulfills them for the $\rm 1.4~M_{\odot}$ configuration, imposed from the GW170817 event, indicated in Figure~\ref{fig:mass_energy_rotation_gibbs}(b). The exception case can only describe the 709 Hz radius at $\rm 1.4~M_{\odot}$ configuration. We also present the corresponding Figures~\ref{fig:mass_density_rotation_maxwell} and~\ref{fig:mass_density_rotation_gibbs} where we highlight the transition density over the energy jump.

The GC configuration presented an effect where while the non-rotating configuration is not able to produce twin stars, that is not the case at 709 Hz~\cite{Banik-2004}. Specifically, Figure~\ref{fig:mr_gibbs_rot} displays the representative cases: (a) $n_{\rm tr} = 0.25~{\rm fm^{-3}}$ and $\Delta \mathcal{E} = \Delta \mathcal{E}_{\rm G} + 120~{\rm MeV~fm^{-3}}$ and (b) $n_{\rm tr} = 0.3~{\rm fm^{-3}}$ and $\Delta \mathcal{E} = \Delta \mathcal{E}_{\rm G} + 60~{\rm MeV~fm^{-3}}$. The last case, $n_{\rm tr} = 0.25~{\rm fm^{-3}}$ and $\Delta \mathcal{E} = \Delta \mathcal{E}_{\rm G} + 90~{\rm MeV~fm^{-3}}$, is not presented for the clearness of the figure. Moreover, for clarity, we indicate in the inner figure, the region of an EOS at 709 Hz, where the twin star branch appears. Also, we have marked the twin star that corresponds to the first maximum mass configuration along with the relevant radii. This effect added the reported cases in the list where the rotation meets the mass limit of the PSR J0952-0607 pulsar.

The MC and GC configurations are both suitable methods to describe twin stars, where the second one, using similar parametrization, is more elaborated. The key differences between them are that (a) the MC provides higher gravitational masses than the GC, and (b) the MC rises fairly more twin stars than the GC. A possible explanation for both of them may lie to the existence of the sharp transition and the energy jump applicable on the MC, where in the GC, both of them are extinct. However, the GC can provide twin stars in the rotating configuration without their appearance in the non-rotating scenario. This peculiar effect may have its origin at the smooth phase transition that the GC uses, alongside with the transition density, the rotational frequency, and the hadronic EOS. Nevertheless, a thorough study for the latter effect should be done and will be the subject of a forthcoming paper. The analysis of Figures~\ref{fig:mass_energy_rotation_maxwell}-\ref{fig:mr_gibbs_rot} enrich the scenario of twin stars existence as they can describe the observed neutron star masses both at non-rotating and rotating configurations. For the numerical integration of the equilibrium equations at the rotating configuration, we used the publicly available numerical code nrotstar from the C++ Lorene/Nrotstar library~\cite{Lorene-1998}.

\section{Concluding Remarks} \label{sec:Remarks}
The existence  of two stable branches in a neutron star configuration,  which lead to the concept of twin stars, may reveal the scenario of phase transition in the interior of dense nuclear matter. Therefore, it is extremely important to study the above hypothesis both from theoretical and  and observant point of view. In the present work a systematic study, concerning the twin star hypothesis,   has been carried out and the main conclusions could be summarized as follows:
\begin{enumerate}
    \item There could be hybrid or twin star branches that can describe the current observations of binary neutron star systems, such as the GW170817 event. Not only the transition density has an important role on the possibility of the existence of a twin star branch, but also the $\mathrm{\Delta\mathcal{E}}$ affects this scenario. Especially, as we move to higher values of $\mathrm{\Delta\mathcal{E}}$, the twin star branch is more easily to exist even for the lowest values of $n_{\rm tr}$ that we used in our study.
    
    \item Across the same EOS, the kind of construction that we use affects the behavior of the curves. In general, the MC offers more hybrid and twin star branches in lower neutron star masses, compared to the GC. Moreover, the MC leads to a softer behavior of the EOS. Therefore, binary neutron star systems that contain lower component masses (otherwise, lower chirp mass), suit better to the MC. Also, the GC shifts the curves to higher values of $\Lambda$, as one can see from the $\Lambda_{1.4}-n_{tr}$ diagram. This fact has as a consequence that some cases which were to be excluded by using the MC, could lie (by using the GC)  as a whole or partly inside the estimated area.
    
    \item In our study we used the observational value of $\tilde{\Lambda}$ using as a reference the GW170817 event. This upper limit was not helpful to constrain further the cases that we studied. On the contrary, because of the fact that the HS-HS binary system scenario has a very soft behavior on the EOS, a possible lower limit on $\tilde{\Lambda}$ could be very helpful to shed more light on the problem, from another point of view. Moreover, a further constraint on the radius (e.g. the radius of a $1.4M_\odot$) could be informative in order to impose constraints on the very soft cases.
    
    \item The observation of the PSR J0952-0607 pulsar, where both mass and frequency are evident, does not exclude the existence of twin stars. In fact, both the MC and the GC provide numerous configurations that can describe solely the rotating configuration, solely the non-rotating configuration, or both of them simultaneously. The latter, since we do not have the corresponding non-rotating mass of the PSR J0952-0607 pulsar, provide us with strong evidence that twin stars can exist as both static and rotating.
    
    \item The rotational frequency, in some cases of the GC, is able to introduce twin stars from normal stars. In particular, as the rotating twin star loses its rotational frequency, the neutron star moves to higher energies and stable configurations, where the EOS is more stiff and extincts the phenomenon of twin stars. However, an analysis using fixed baryon mass sequences would provide more information for the effect. Nevertheless, we speculate that this effect due to the stiffening of the EOS, comes as a consequence to the GC, where a smooth phase transition is employed, in contrast to the MC, where an energy jump appears.
    
    \item The recent observation of a compact object with the lowest observable  mass $M=0.77_{-0.17}^{+0.20}\ M_{\odot}$,  within the supernova remnant HESS J1731-347, opens a new window to impose constraints on the EOS of dense nuclear matter. In the present study, our predictions reinforce the estimation of the authors in Ref.~\cite{Doroshenko-2022} that the observed compact object is rather a hybrid star with exotic core than the lightest neutron star ever known. This result strengthens the hypothesis of hybrid stars and therefore of the existence of twin stars.
    
    \end{enumerate}
    
The above conclusions  can be summarized as follows: Further systematic theoretical research is required to clarify the role of the method that describes the phase transition in dense nuclear matter, as well as its particular characteristics (density transition, energy, etc.). In addition, more relevant observations from binary neutron star merger (and not only)  are necessary to be able to check the plausibility of the theoretical predictions. Possibly, in this case we will be able to confirm, with enough confidence, the existence of twin stars but even more importantly, to confirm, not only qualitatively but also quantitatively, the phenomenon of phase transition in dense nuclear matter.

\section*{Acknowledgements}
The authors would like to thank Prof. D. Blaschke, Prof. V. Dexheimer, and Prof. K. Kokkotas for their useful insight and comments and Dr. S. Typel for useful discussions and correspondence. The research was supported, in part, by the Deutsche Forschungsgemeinschaft (DFG) through the cluster of excellence ct.qmat (Exzellenzcluster 2147, project-id 390858490) and by the Hellenic Foundation for Research and Innovation (HFRI) under the 3rd Call for HFRI PhD Fellowships (Fellowship Number: 5657). The implementation of the research was co-financed by Greece and the European Union (European Social Fund-ESF) through the Operational Programme $\ll$Human Resources Development, Education and Lifelong Learning$\gg$ in the context of the Act ``Enhancing Human Resources Research Potential by undertaking a Doctoral Research" Sub-action 2: IKY Scholarship Programme for PhD candidates in the Greek Universities.



\begin{thebibliography}{32}
	%
	\bibitem{Glendenning-1997} N. K. Glendenning, Compact stars - Nuclear physics, Particle physics and General Relativity, Springer (1997).
	%
    \bibitem{Haensel-2007} P. Haensel, A. Y. Potekhin, and D. G. Yakovlev, Neutron Stars 1: Equation of State and Structure (Springer-Verlag, New York, 2007).
    %
    \bibitem{Bielich-2020} J. Schaffner-Bielich, Compact Star Physics (Cambridge University Press, Cambridge, England,2020).
    %
    \bibitem{Weber-1999}] F. Weber, Pulsars as Astrophysical Laboratories for Nuclear and Particle Physics (Institute of Physics, Bristol, U.K., 1999).
    %
    \bibitem{Abbott-1} B.P. Abbott {\it et al.}, Phys. Rev. Lett. \textbf{119}, 161101 (2017).
    %
    \bibitem{Abbott-2} B. P. Abbott {\it et al.}, Phys. Rev. Lett. \textbf{121}, 161101 (2018).
    %
    \bibitem{Abbott-gw170817} B. P. Abbott {\it et al.}, Phys. Rev. X \textbf{9}, 011001 (2019).
    %
    \bibitem{Baym-2018}G. Baym, T. Hatsuda, T. Kojo, P.D. Powell, Y. Song, and T. Takatsuka, Rep. Prog. Phys. {\bf 81} 056902 (2018). 
    \bibitem{Alford-2013}M.G. Alford, S. Han, and M. Prakash, Phys. Rev. D \textbf{88}, 083013 (2013).
    %
	\bibitem{Christian-2018} J.E. Christian, A. Zacchi, and  J. Schaffner-Bielich, Eur. Phys. J. A \textbf{54}, 28 (2018).
	%
	\bibitem{Montana-2019} G. Montana, L. Tolos, M. Hanauske, L. Rezzolla, Phys. Rev. D \textbf{99}, 103009 (2019).
	%
    \bibitem{Gerlach-1968}	U.H. Gerlach, Phys. Rev. {\bf 172}, 1325 (1968).
    %
    \bibitem{Kampfer-1981a}Kämpfer, J. Phys. A: Math. Gen. {\bf 14}, L471, (1981).
    %
    \bibitem{Kamfer-1981b}Kämpfer, Phys. Lett.  B {\bf 101}, 366 (1981).
	%
	\bibitem{Glendenning-2000} N. K. Glendenning and C. Kettner, Astron. Astrophys. \textbf{353}, 811 L9-L12 (2000).
	%
	\bibitem{Schertler-2000}K. Schertler, C. Greiner, J. Schaffner-Bielich, and M.H. Thoma, Nucl. Phys. A {\bf 677}, 463 (2000).
    %
    \bibitem{Blaschke-2013}D. Blaschke, D.E. Alvarez-Castillo, and S. Benic, PoS CPOD2013 063 (2013); [arXiv:1310.3803].
    %
    \bibitem{Castillo-2013}D.E. Alvarez-Castillo  and  D. Blaschke, arXiv:1304.7758.
    %
    \bibitem{Benic-2015}S. Benic, D. Blaschke, D.E. Alvarez-Castillo, T. Fischer, and S. Typel, Astron. Astrophys. {\bf 577}, A40 (2015).
    %
    \bibitem{Castillo-2015}D.E. Alvarez-Castillo  and  D. Blaschke,  Phys. Part. Nucl. {\bf 46}, 846 (2015).
    %
    \bibitem{Castillo-2016}D.E. Alvarez-Castillo, A. Ayriyan, S. Benic, D. Blaschke. H. Grigorian, and S. Typel, Eur. Phys. J. A {\bf 52}, 69(2016).
    %
    \bibitem{Ayriyan-2018}A. Ayriyan, N.-U. Bastian, D. Blaschke, H. Grigorian, K. Maslov, and D. N. Voskresensky, Phys. Rev. C \textbf{97}, 045802 (2018).
    %
    \bibitem{Maslov-2019} K. Maslov, N. Yasutake, D. Blaschke, A. Ayriyan, H. Grigorian, and T. Maruyama, Phys. Rev. C \textbf{100}, 025802 (2019).
    %
    %
	\bibitem{Alvarez-2017}   D. E. Alvarez-Castillo and D. B. Blaschke, Phys. Rev. C \textbf{96}, 045809, (2017)
	%
	\bibitem{Bejger-2017} M. Bejger,  D. Blaschke, P. Haensel, J.L. Zdunik, and M. Fortin, Astron. Astrophys. \textbf{600}, A39 (2017).
	%
	\bibitem{Bhattacharyya-2010} A. Bhattacharyya, I. N. Mishustin, W. Greiner, J. Phys. G \textbf{37}, 025201 (2010).
	%
	%
	\bibitem{Christian-2019} J.E. Christian, A. Zacchi, and  J. Schaffner-Bielich, Phys. Rev. D \textbf{99}, 023009 (2019).
	%
	\bibitem{Christian-2021} J.E. Christian, and  J. Schaffner-Bielich, Phys. Rev. D \textbf{103}, 063042. (2021)
	%
	\bibitem{Christian-2022} J.E. Christian, and  J. Schaffner-Bielich, Astrophys. J. {\bf 935}, 122 (2022).
	%
	\bibitem{Han-2019a} S. Han and A.W. Steiner, Phys. Rev. D \textbf{99}, 083014 (2019).
	%
	\bibitem{Espinoza-2022}P. L. Espino and V. Paschalidis, Phys. Rev. D {\bf 105}, 043014 (2022).
	%
	\bibitem{Tan-2022}H. Tan, T. Dore, V. Dexheimer, J. Noronha-Hostler, and N. Yunes, Phys. Rev. D {\bf 105}, 023018 (2022).
	%
	\bibitem{Li-2021}J. J. Li, A. Sedrakian, and M. Alford, Phys. Rev. D {\bf 104}, L121302 (2021).
	%
	\bibitem{Sharifi-2021}Z. Sharifi, M. Bigdeli, and D. Alvarez-Castillo, Phys. Rev. D {\bf 103}, 103011 (2021).
	%
	\bibitem{Benitez-2021}E. Benitez, J. Weller, V. Guedes, C. Chirenti, and M. Coleman Miller, Phys. Rev. D {\bf 103}, 023007 (2021). 
	%
	\bibitem{Sendra-2020}J.J. Li, A. Sedrakian, and M. Alford, Phys. Rev. D {\bf 101}, 063022 (2020).
	%
	\bibitem{Castillo-2019} D.E. Alvarez-Castillo, D.B. Blaschke, A.G. Grunfeld, and V.P. Pagura, Phys. Rev. D {\bf 99}, 063010 (2019). 
	%
	\bibitem{Paschalidis-2018} V. Paschalidis, K. Yagi, D. Alvarez-Castillo, D.B. Blaschke, and A. Sedrakian, Phys. Rev. D {\bf 97}, 084038 (2018).
	%
	\bibitem{Spinella-2017}I.F. Ranea-Sandoval, M.G. Orsaria, S. Han, F. Weber, and W.M. Spinella, Phys. Rev. C {\bf 96}, 065807 (2017). 
    %
	\bibitem{Alford-2017}M. Alford and A. Sedrakian, Phys. Rev. Lett. {\bf 119}, 161104 (2017).
    %
	\bibitem{Zacchi-2017}A. Zacchi, L. Tolos, and J. Schaffner-Bielich, Phys. Rev. D {\bf 95}, 103008 (2017). 
    %
    \bibitem{Zacchi-2016}A. Zacchi, M. Hanauske, and J. Schaffner-Bielich, Phys. Rev. D {\bf 93}, 065011 (2016).
    %
    \bibitem{Bhattacharyya-2005}A. Bhattacharyya, S. K. Ghosh, M. Hanauske, and S. Raha, Phys. Rev. C {\bf 71}, 048801 (2005). 
    %
   \bibitem{Sen-2022}D. Sen, N. Alam, and G. Chaudhuri, Phys. Rev. D {\bf 106}, 083008 (2022).
    %
   \bibitem{Minamikawa-2021}T. Minamikawa, T. Kojo, and M. Harada, Phys. Rev. C {\bf 103}, 045205 (2021). 
    %
	\bibitem{Pietri-2019}R. De Pietri, A. Drago, A. Feo, G. Pagliara, M. Pasquali, S. Traversi, and G. Wiktorowicz, Astrophys. J. {\bf 881}, 122 (2019).
    %
    \bibitem{Zdunik-2013}J.L. Zdunik  and P. Haensel, Astron. Astrophys. \textbf{551}, A61 (2013).
	%
	\bibitem{Han-2020}S. Han and M. Prakash, Astrophys. J. {\bf 899}, 164 (2020).
    %
	\bibitem{Li-2020} A. Li, Z.Y. Zhu, E.P. Zhou, J.M. Dong, J.N. Hu, C.J. Xia, Journ. High Energ. Astroph. {\bf 28}, 19 (2020).
	%
	\bibitem{Largani-2022} N.K. Largani, T. Fischer, A. Sedrakian, M. Cierniak, D.E. Alvarez-Castillo, and  D. B. Blaschke, Mon. Not. Roy. Astron. Soc. {\bf 515}, 3539 (2022).
    %
	\bibitem{Ivanytskyi-2022}O. Ivanytskyi and D. Blaschke, Phys. Rev. D {\bf 105}, 114042 (2022).
	%
	\bibitem{Contrera-2022}G.A. Contrera, D. Blaschke, J.P. Carlomagno, A.G. Grunfeld, and S. Liebing, Phys. Rev. C {\bf 105}, 045808 (2022). 
	%
	\bibitem{Schram-2016}S. Schramm, V. Dexheimer, and  R. Negreiros, Eur. Phys. J. A \textbf{52}, 14 (2016). 
	%
	\bibitem{Burgio-2018} G.F. Burgio, A. Drago, G. Pagliara, H.J. Schulze, and J.B. Wei, Astrophys. J. {\bf 860}, 139 (2018).
    %
   \bibitem{Sieniawska-2019} M. Sieniawska, W. Turczanski, M. Bejger, and J.L. Zdunik, Astron. Astrophys. {\bf 622}, A174 (2019).
    %
    \bibitem{Most-2018} E.R. Most, L.R. Weih, L. Rezzolla, and J. SchaffnerBielich, Phys. Rev. Lett. {\bf 120}, 261103 (2018).
    %
    \bibitem{Nandi-2018}  R. Nandi and P. Char, Astrophys. J. {\bf 857}, 12 (2018). 	
	%
    \bibitem{Deloudis-2021}	T. Deloudis, P. Koliogiannis, Ch. Moustakidis, EPJ Web Conf. {\bf 252}, 06001 (2021).
    %
    \bibitem{Wang-2022}Q.W. Wang, C. Shi, Y. Yan, and  H.S. Zong, Nucl. Phys. A {\bf 1025}, 122489 (2022). 
    %
    \bibitem{Banik-2004} S. Banik, M. Hanauske, D. Bandyopadhyay, and W. Greiner, Phys. Rev. D \textbf{70}, 123004 (2004).
    %
    \bibitem{Banik-2001}S. Banik and D. Bandyopadhyay, Phys. Rev. D \textbf{64}, 055805 (2001).
    \bibitem{Banik-2003}S. Banik and D. Bandyopadhyay, Phys. Rev. D \textbf{67}, 123003 (2003).
    \bibitem{Haensel-2016}P. Haensel, M. Bejger, M. Fortin, and L. Zdunik, Eur. Phys. J. A \textbf{52}, 59 (2016).
    %
        %
    \bibitem{Bozzola-2019} G. Bozzola, P.L. Espino, C.D. Lewin, and V. Paschalidis, Europ. Phys. J. A \textbf{55}, 149 (2019).
	%
	\bibitem{Pereira-2020} J.P. Pereira, M. Bejger, N. Andersson, and F. Gittins, Astrophys. J. {\bf 895}, 28 (2020).
	
	\bibitem{Han-2019} S. Han, M.A.A. Mamun, S. Lalit, C. Constantinou, M. Prakash, Phys. Rev. D \textbf{100}, 103022 (2019).
	
	\bibitem{Bassa-2017}C.G. Bassa, Z. Pleunis, J.W.T. Hessels, {\it et al.}, Astrophys. J. Lett. {\bf 846}, L20 (2017).
    %
    \bibitem{Romani-2022}R.G. Romani, D. Kandel, A.V. Filippenko, T.G. Brink, W. Zheng, Astrophys. J. Lett. {\bf 934}, L17 (2022).
	%
	\bibitem{Ecker-2022}C. Ecker and L. Rezzolla, arXiv:2209.08101. 
	\bibitem{Doroshenko-2022}V. Doroshenko, V. Suleimanov, G. Pühlhofer, and  Andrea Santangelo, Natur. Astr. (2022); https://doi.org/10.1038/s41550-022-01800-1.
     \bibitem{Mariani-2017} M. Mariani, M. Orsaria,  and H. Vucetich, Astron. Astrophys. {\bf 601 }  A21 (2017).  

	\bibitem{Margaritis-2020} Ch. Margaritis, P.S. Koliogiannis, and Ch.C. Moustakidis, Phys. Rev. D {\bf 101}, 043023 (2020).
	\bibitem{Kanakis-2020} A. Kanakis-Pegios, P.S. Koliogiannis, and Ch.C. Moustakidis, Phys. Rev. C {\bf 102}, 055801 (2020).
	\bibitem{Koliogiannis-2021f} P.S. Koliogiannis, A. Kanakis-Pegios, and Ch.C. Moustakidis, Foundations {\bf 1(2)}, 217-255 (2021).
	
	\bibitem{Seidov-1971} Z.F. Seidov, Soviet Astronomy, \textbf{15}, 347 (1971)
	%
	\bibitem{Lighthill-1950} J.M. Lighthill, Mon. Not. Roy. Astron. Soc., \textbf{110}, 339-342 (1950).
	%
    \bibitem{Postnikov-2010} S. Postnikov, M. Prakash, J. M. Lattimer, Phys. Rev. D {\bf82}, 024016 (2010).
    %
    \bibitem{Flanagan-08} E.E. Flanagan, T. Hinderer, Phys. Rev. D {\bf77}, 021502(R) (2008).
    %
    \bibitem{Hinderer-08} T. Hinderer, Astrophys. J. {\bf677}, 1216 (2008).
    %
    \bibitem{Hinderer-10} T. Hinderer, B.D. Lackey, R. N. Lang, J.S. Read, Phys.Rev. D {\bf81}, 123016 (2010).
    %
    \bibitem{Koliogiannis-2021} P.S. Koliogiannis and Ch.C. Moustakidis, Astrophys. J. \textbf{912}, 69 (2021).
    %
    \bibitem{Typel-2018} S. Typel, J. Phys. G: Nucl. Part. Phys. \textbf{45}, 114001 (17pp) (2018).
    %
    
    
    %
	%
	%
	%
	%
	%
	\bibitem{Arzoumanian-2018} Z. Arzoumanian, A. Brazier, S. Burke-Spolaor, et al., Astrophys. J. Suppl. S. 
	%
	\bibitem{Antoniadis-2013} J. Antoniadis, P. Freire, N. Wex, et al., Sci. \textbf{340}, 448 (2013).
	%
	\bibitem{Cromartie-2020} H. Cromartie, E. Fonseca, S. Ransom, et al., Nat. Astron. \textbf{4}, 72 (2020).
	%
	%
	\bibitem{Capano-2020} C.D. Capano, I. Tews, S.M. Brown, B. Margalit, S. De, S. Kumar, D.A. Brown, B. Krishnan, and S. Reddy, Nat. Astron. \textbf{4}, 625-632 (2020).
	%
	\bibitem{Lorene-1998} LORENE 1998, LORENE: Langage Objet pour la RElativité NumériquE, http://lorene.obspm.fr/
\end{thebibliography}
\end{document}